\documentclass[%
reprint,
superscriptaddress,
amsmath,amssymb,amsthm,mathrsfs,
aps,
prl,
]{revtex4-2}

\usepackage{graphicx}
\usepackage{dcolumn}
\usepackage{bm}
\usepackage[usenames]{color}
\usepackage[ 
margin=0.75in,
]{geometry}
\usepackage[normalem]{ulem}

\graphicspath{{./figs/}}

\begin{document}
	
\title{
	Giant spontaneous magnetostriction in MnTe driven by a novel magnetostructural coupling mechanism
}
	
	\author{Raju Baral}
	\affiliation{ %
		Department of Physics and Astronomy, Brigham Young University, Provo, Utah 84602, USA.
	} %

	\author{Milinda Abeykoon}
	\affiliation{ %
		Photon Sciences Division, Brookhaven National Laboratory, Upton, NY, 11973 USA.
	} %
 
	\author{Branton J. Campbell}
	\affiliation{ %
		Department of Physics and Astronomy, Brigham Young University, Provo, Utah 84602, USA.
	} %

	\author{Benjamin A. Frandsen}
	\affiliation{ %
		Department of Physics and Astronomy, Brigham Young University, Provo, Utah 84602, USA.
	} %

\begin{abstract}
We present a comprehensive x-ray scattering study of spontaneous magnetostriction in hexagonal MnTe, an antiferromagnetic semiconductor with a N\'eel temperature of $T_{\mathrm{N}} = 307$~K. We observe the largest spontaneous magnetovolume effect known for an antiferromagnet, reaching a volume contraction of $|\Delta V/V| > 7 \times 10^{-3}$. This can be justified semiquantitatively by considering bulk material properties, the spatial dependence of the superexchange interaction, and the geometrical arrangement of magnetic moments in MnTe. The highly unusual \textit{linear} scaling of the magnetovolume effect with the short-range magnetic correlations, beginning in the paramagnetic state well above $T_{\mathrm{N}}$, points to a novel physical mechanism, which we explain in terms of a trilinear coupling of the elastic strain with superposed distinct domains of the antiferromagnetic order parameter. This novel mechanism for coupling lattice strain to robust short-range magnetic order casts new light on magnetostrictive phenomena and also provides a template by which the exceptional magnetostrictive properties of MnTe might be realized in a wide range of other functional materials.

\end{abstract}
	
\maketitle

\section{Introduction}

Despite its simple composition and structure, hexagonal manganese telluride (MnTe; space group $P6_3/mmc$, \#194~\cite{johns;jinc61}) exhibits a rich variety of electronic and magnetic behaviors that have sparked a surge of recent interest and have shown promise in numerous technological applications. As an antiferromagnetic (AFM) semiconductor with a N\'eel temperature of 307~K, MnTe has emerged as a promising platform for AFM spintronics~\cite{krieg;nc16,yin;prl19,bossi;prb21}. When doped with charge carriers, it demonstrates outstanding thermoelectric properties~\cite{xu;jmchema17,ren;jmchemc17,dong;jmchemc18,deng;nanoe21,xiong;sma23,zulki;advs23}, due in large part to the influence of short-range magnetic correlations on the thermopower at high temperature~\cite{zheng;sadv19,polas;crps21,baral;matter22}. MnTe can also be viewed as a structural component of the intrinsic magnetic topological insulator MnBi$_2$Te$_4$, in which Bi$_2$Te$_3$ is intercalated by layers of magnetic MnTe~\cite{deng;s20}, extending the interest in MnTe into the realm of topological quantum materials. More philosophically, MnTe has intrigued researchers as a ``crossroads'' material for its electronic properties, which are intermediate between strongly correlated insulators like MnO and more weakly correlated metals like MnSb~\cite{allen;ssc77, youn;pssb04}.

Yet another promising aspect of MnTe, although one that has received little attention to date, is the magnetostructural behavior associated with the AFM transition. In this work, we explore the phenomenon of magnetostriction, by which the dimensions of a solid change in response to an applied magnetic field (forced magnetostriction) or the development of intrinsic magnetic order (spontaneous magnetostriction). This effect has found widespread application in both established and emerging technologies~\cite{dutre;b;mtaaom93}, including sensors, actuators, and transducers~\cite{hrist;jmmm07}; biomedical devices~\cite{gao;bam22}; SONAR~\cite{angus;u69,jile;jpd94}; zero- or negative-thermal-expansion materials~\cite{song;pms21}; and straintronics for energy efficient computing~\cite{bandy;aprev21}, among others. However, a comprehensive understanding of the magnetostrictive effects observed in many classes of magnetic materials is lacking, particularly for spontaneous magnetostriction~\cite{sande;b;hmmm20}. Although studies of magnetostriction have historically focused on ferromagnetic materials, the tremendous diversity of antiferromagnetic phases provides an outstanding opportunity for fundamental investigations of magnetostriction~\cite{doerr;advp05}, leading to a burgeoning interest in magnetostrictive antiferromagnets~\cite{chari;prb12,singh;npjcm21,miao;prb21,casil;prm21,dey;prb21}.


Here, we use x-ray scattering to gain detailed insight into the spontaneous magnetostriction in MnTe in response to the AFM transition. We observe a magnetically-driven volume contraction greater than 0.7\% in MnTe and Na-doped MnTe, which constitutes the largest known magnetovolume response in an antiferromagnet by a wide margin and is comparable in magnitude to values observed in commercially exploited ferromagnets exhibiting ``giant'' magnetostriction such as Terfenol-D~\cite{jile;jpd94}. The magnetostructural response is initially driven by local magnetic correlations that develop well above the N\'eel temperature, highlighting the influence of short-range correlations on macroscopic properties. Surprisingly, the lattice contraction scales linearly with the local magnetic order parameter $m$, in contrast to the typical $m^2$ dependence that is widely observed in other materials. This linear response points to a previously unknown magnetostrictive mechanism, which we show can be explained by free energy arguments that explicitly consider the coupling of distinct domains of a short-range magnetic order parameter. In addition to revealing the exceptional magnetostructural properties of MnTe, this work provides new insights into fundamental aspects of spontaneous magnetostriction.

\section{Experimental details}
Polycrystalline samples of pure MnTe and 2\% Na-doped MnTe were synthesized as described in Ref.~\onlinecite{baral;matter22}. X-ray scattering experiments were performed at beamline 28-ID-1 of the National Synchroton Light Source II (NSLS-II) at Brookhaven National Laboratory. The samples were loaded into polyimide capillaries and placed in a liquid helium cryostat mounted on the beamline. Temperature-dependent data were collected using a wavelength of 0.19~\AA. Additional measurements were performed at selected temperatures with an \textit{in situ} magnetic field using a 5-T magnet mounted on the beamline. The wavelength was 0.106~\AA\ for the field-dependent measurements. The diffraction patterns were azimuthally integrated using DIOPTAS~\cite{presc;hpr15} to obtain one-dimensional intensity patterns, which were then normalized and Fourier transformed in XPDFsuite~\cite{yang;arxiv15} to produce atomic pair distribution function (PDF) data, which is sensitive to both the average and local structure~\cite{egami;b;utbp12}. The maximum momentum transfer included in the Fourier transform was 25~\AA$^{-1}$ for the temperature-dependent data and 18~\AA$^{-1}$ for the field-dependent data. Structural fits to the data using the published hexagonal crystal structure of MnTe were performed with PDFgui~\cite{farro;jpcm07} over the real-space range of 1.5 -- 50~\AA. Representative fits are shown in the Supplementary Information (SI). 
We performed symmetry analyses of the magnetic order parameters and the invariant free energy polynomial using the ISODISTORT~\cite{campb;jac06,stoke;webIso} and INVARIANTS~\cite{hatch;jac03,stoke;webInv} programs of the ISOTROPY Software Suite (\url{https://iso.byu.edu}), respectively.

\section{Results}
\begin{figure}
	\centerline{\includegraphics[ width=75mm]{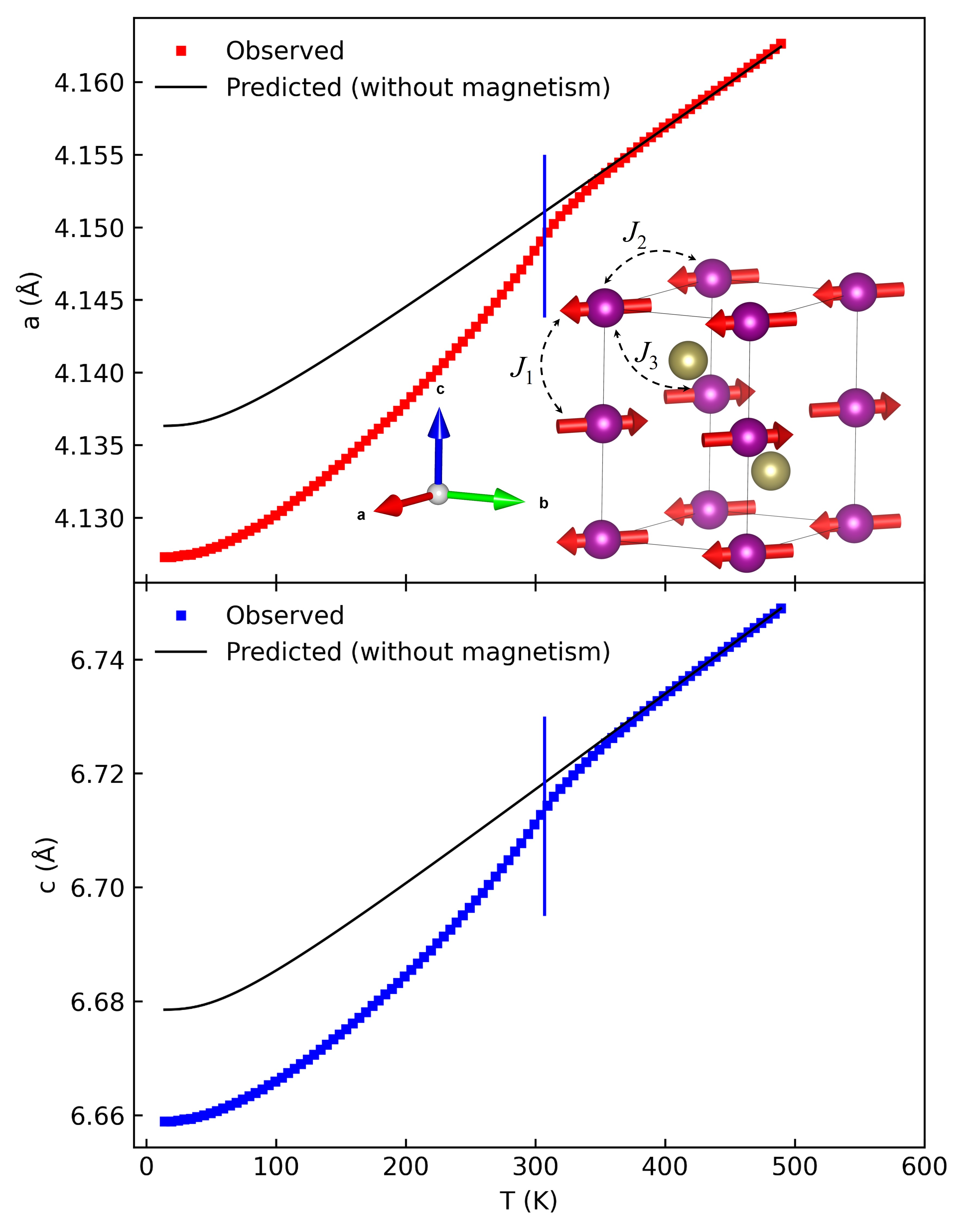}}
	\caption{\label{fig:MnTe_debyefit} (a) Temperature dependence of the observed values of $a$ and the prediction based on the Debye-Gr\"uneisen model, which does not include magnetism. Error bars are smaller than the symbol size. Inset: Crystal and magnetic structure of MnTe with the first three magnetic exchange interactions labeled as $J_1, J_2,$ and $J_3$. Purple (gold) spheres represent Mn (Te) atoms. The red arrows represent the Mn$^{2+}$ magnetic moments. (b) Same as panel (a), but for $c$.}		
\end{figure}

The hexagonal lattice parameters \textit{a} and \textit{c} are shown in Fig.~\ref{fig:MnTe_debyefit} (a) and (b) as a function of temperature as determined from fits to the x-ray PDF data.
With decreasing temperature from 500 to $\sim$375~K, the lattice parameters decrease linearly, as expected for conventional thermal contraction. However, upon further cooling toward $T_{\mathrm{N}} \sim$~307 K, the lattice parameters decrease at a significantly faster rate, dropping below the trend extrapolated from higher temperature. We observe a similar effect in Na-doped MnTe (see SI). This is clear evidence for spontaneous magnetostriction in MnTe in response to the development of magnetic order.

To probe this behavior more quantitatively, we modeled the expected temperature dependence of the lattice parameters and unit cell volume in the absence of magnetism using the Gr\"uneisen equation of state combined with the Debye approximation of the internal energy. According to this Debye-Gr\"uneisen model, the unit cell volume as a function of temperature $T$ is given by~\cite{chatt;jpcm10,chatt;jpcm11}
\begin{equation}
\label{eq:gruneisen}
    V(T) =\alpha T\left(\frac{T}{T_{D}}\right)^3  \int_{0}^{T_{D}/T} \frac{x^3}{(e^{x} - 1)} \,dx + V_0 ,
\end{equation}
where $V(T)$ is the temperature-dependent unit cell volume, $V_0$ is the volume at 0~K, $\alpha$ is given by  $9 N k_{B} \gamma/B_{0}$, $\gamma$ is the Gr\"uneisen parameter, $N=4$ is the number of atoms per unit cell, $B_0$ is the bulk modulus at 0~K, and $T_{\mathrm{D}}$ is the Debye temperature, determined by previous experiments to be 223~K~\cite{zheng;sadv19}. The cubed root of the right hand side of Eq.~\ref{eq:gruneisen} can be used for the individual lattice parameters $a$ and $c$~\cite{chatt;jpcm10,chatt;jpcm11}. Fitting this model to the lattice parameter and unit cell volume data above 420~K, we can determine values for $\gamma/B_{0}$ and $V_{0}$, which we then use to calculate the full temperature dependence down to 0~K. The best-fit values of $\gamma/B_0$ (see SI) are close to the values determined for similar materials such as MnF$_2$~\cite{chatt;jpcm10} and consistent with the reported bulk modulus of MnTe~\cite{paszk;appa97}. We note that using a wide range of Debye temperatures between 140~K and 300~K makes only a small difference to the fitting results. An equivalent analysis was also performed for the Na-doped sample (see SI). 

\begin{figure}
	\centerline{\includegraphics[width=80mm]{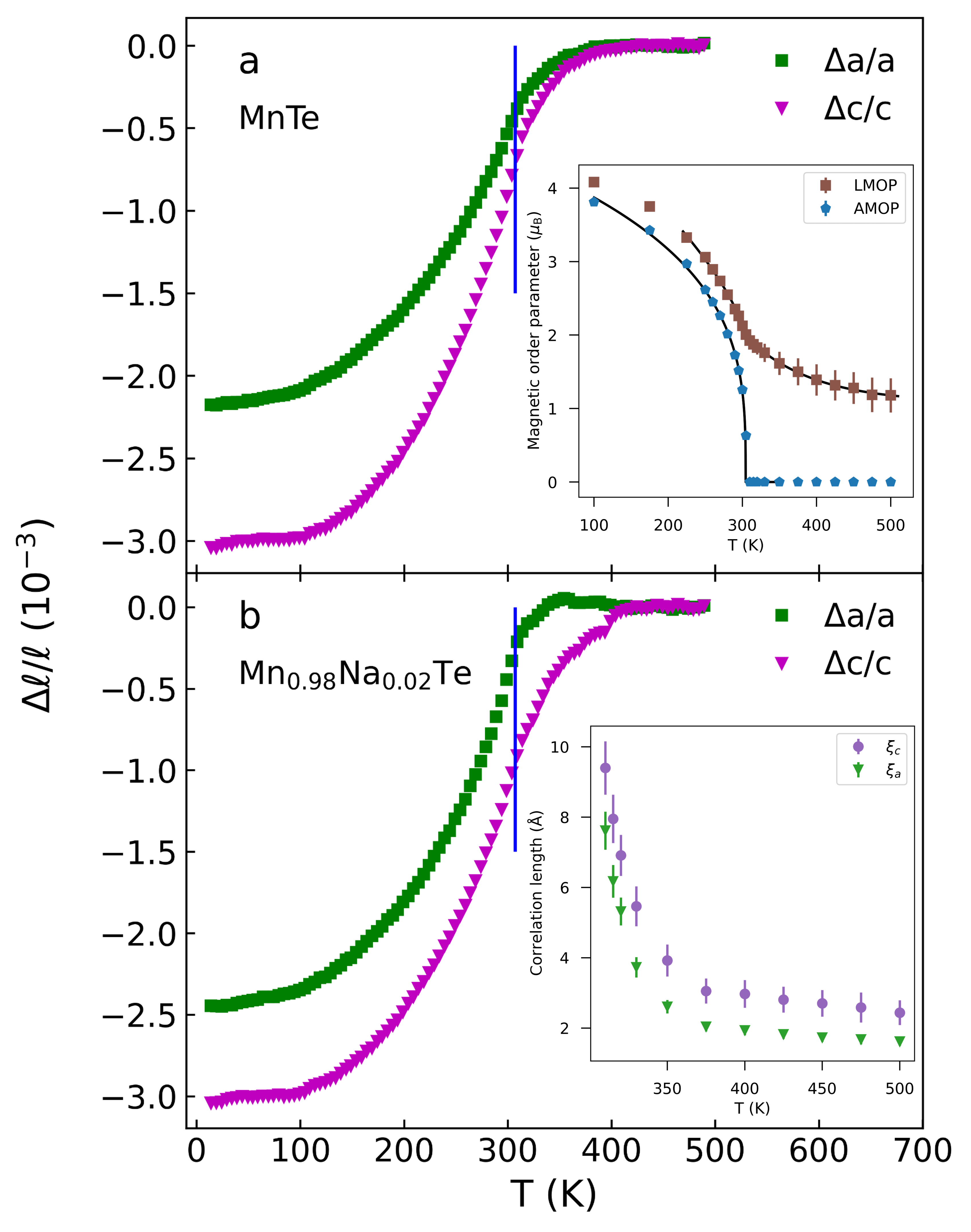}}
	\caption{\label{fig:fractionalchange} (a) Spontaneous magnetostriction along the $a$ and $c$ axes in pure MnTe. Inset: Local and average magnetic order parameter (LMOP and AMOP) as a function of temperature. (b) Same as panel (a), but for Na-doped MnTe. Inset: Temperature dependence of the magnetic correlation length along $a$ and $c$ in the paramagnetic state.}		
\end{figure}

The calculated lattice parameter trends are shown by the black curves in Fig.~\ref{fig:MnTe_debyefit}. These curves represent the expected temperature dependence of the lattice parameters without any magnetic influence; thus, the difference between the observed and predicted lattice parameter values can be attributed to spontaneous magnetostriction associated with the magnetic phase transition. The magnetostriction is quantified as $\Delta\ell/\ell$, where $\ell$ is the observed lattice parameter (either $a$ or $c$ in our case) and $\Delta\ell$ is the observed lattice parameter minus the predicted value. We plot this value as a function of temperature for both the pure and Na-doped samples in Fig.~\ref{fig:fractionalchange}.
In all cases, $\Delta\ell/\ell$ is close to zero above 400~K, indicating that the calculated lattice parameters accurately capture the high-temperature trend. Between 400 and 350~K, $\Delta\ell/\ell$ begins to trend negative, marking the onset of the spontaneous magnetostriction. This is well above $T_N=307$~K (marked by the vertical blue lines in Fig.~\ref{fig:fractionalchange}), indicating that short-range magnetic correlations in the paramagnetic state initially drive the magnetostriction. To demonstrate this further, we show in the inset to Fig.~\ref{fig:fractionalchange}(a) the average magnetic order parameter (AMOP) and the local magnetic order parameter (LMOP) determined from magnetic PDF analysis~\cite{frand;aca14,frand;aca15}, as recently reported elsewhere~\cite{baral;matter22}. The AMOP corresponds to long-range magnetic order and shows a conventional transition at 307~K, while the LMOP corresponds to nearest-neighbor magnetic correlations and persists deep into the paramagnetic phase.
The inset to Fig.~\ref{fig:fractionalchange}(b) shows the thermal evolution of the in-plane and out-of-plane correlation lengths ($\xi_a$ and $\xi_c$, respectively), which exhibit a strong increase as the temperature is lowered below $\sim$350~K, coinciding with the onset of the magnetostructural response. This is further evidence that the local magnetic correlations in the paramagnetic state drive the spontaneous magnetostriction when the correlation length is sufficiently large. These observations provide useful insights into models of magnetostructural coupling when short-range magnetic correlations are present. 
We expect this coupling between the long-range crystal structure and short-range magnetic correlations to be relevant to numerous other magnetic materials that have a correlated paramagnetic state.

\begin{figure}
	\centerline{\includegraphics[width=70mm]{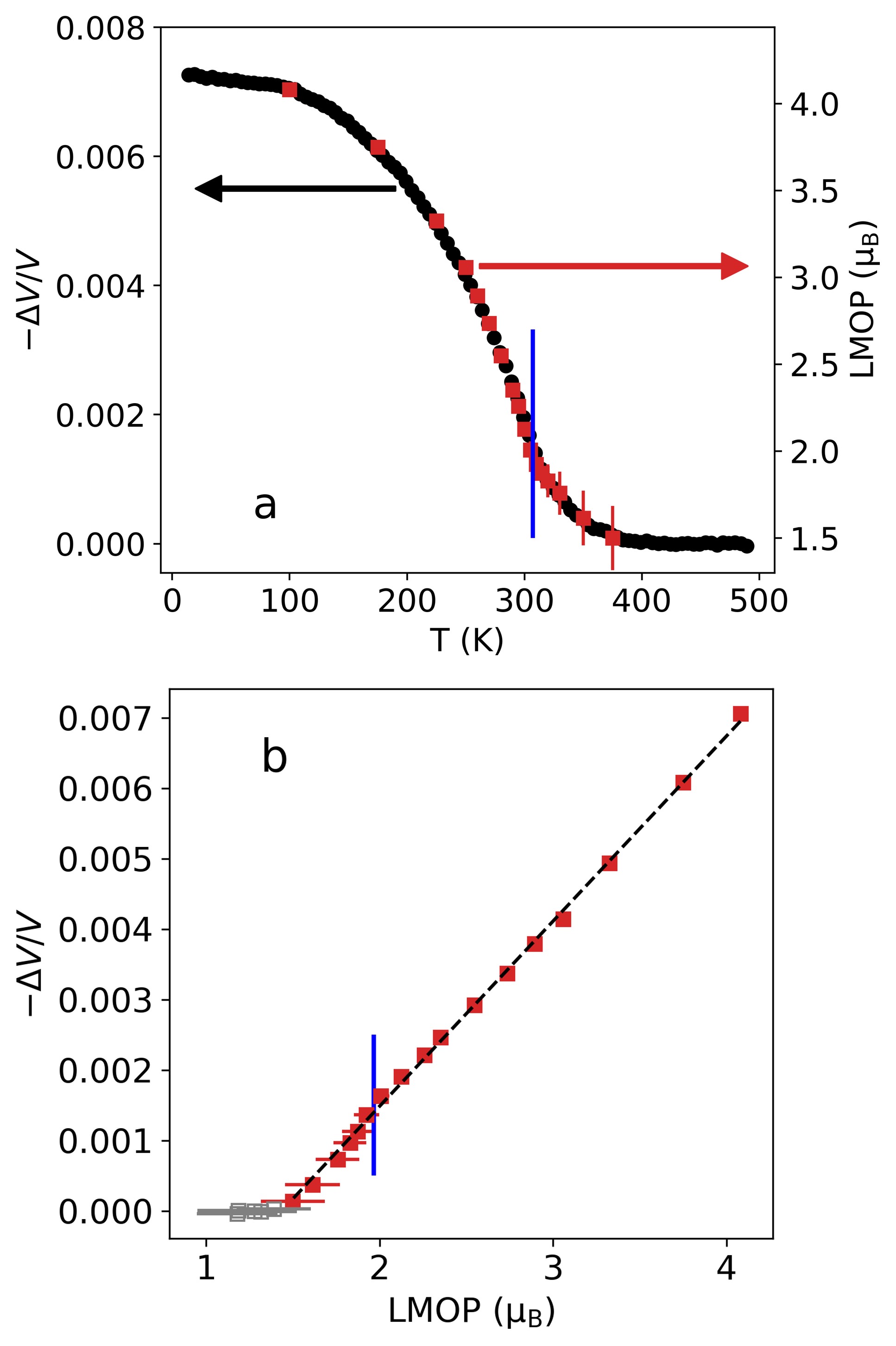}}
	\caption{\label{fig:magnetovolume} (a) Temperature dependence of the magnitude of the magnetovolume shift in MnTe (black symbols, left vertical axis) and the local magnetic order parameter (red symbols, right vertical axis). $T_{\mathrm{N}}$ is marked by the blue vertical line. (b) Magnetovolume shift versus local magnetic order parameter, showing a linear correlation. The blue line marks the long-range ordering transition; points to the left (right) of the line are in the paramagnetic (long-range AFM) state. The dashed line is the best-fit line to the data points shown by red symbols. The gray data points were excluded because they correspond to high temperatures above the onset of the magnetostructural response.}		
\end{figure}

Fig.~\ref{fig:magnetovolume}(a) displays the relative volume change, $-\Delta V/V$, in black symbols, with the minus sign included for convenience to produce an overall positive quantity. 
Upon cooling, the magnetovolume shift begins to deviate from zero slightly below 400~K. At the lowest measured temperature,  $\Delta V/V=-7.3\times 10^{-3}$ for pure MnTe and $-7.8\times 10^{-3}$ for Na-doped MnTe (see the SI). To our knowledge, this is the largest spontaneous magnetovolume effect reported for any antiferromagnetic material to date and is several times larger than that observed in model antiferromagnets like CrF$_{2}$ ($ -7.72 \times 10^{-4}$), CuF$_2$ ($-1.16 \times 10^{-4}$) \cite{chatt;jpcm11}, MnF$_2$ ($-1 \times 10^{-3}$), and NiF$_2$ ($-4.5 \times 10^{-4}$)~\cite{chatt;jpcm10}. Even for ferromagnetic materials, this would be a large magnetovolume effect, reaching a magnitude similar to that observed in commercial magnetostrictive materials like Terfenol-D~\cite{jile;jpd94}. At the same time, MnTe exhibits no observable forced magnetostriction in fields up to 5~T (see the SI), which is unsurprising for an antiferromagnet.

We explore the nature of the magnetostructural coupling in more detail by comparing the magnetic order parameter to the spontaneous magnetovolume shift. The LMOP is plotted as a function of temperature on the right vertical axis in Fig.~\ref{fig:magnetovolume}(a), where we see that the LMOP data points below 400~K fall directly on the magnetovolume curve (left vertical axis) for the selected axis scaling. This reveals a striking direct proportionality between the LMOP and the magnetostriction, as demonstrated in Fig.~\ref{fig:magnetovolume}(b). A linear relationship is evident from the lowest temperature measured (100~K; upper right corner of the plot) up to about 400~K (lower left region of the plot).  Above 400~K, the magnetovolume shift goes to zero while the LMOP persists. The Na-doped sample also exhibits this linear coupling, as shown in the SI. We conclude that in both the paramagnetic and antiferromagnetic phases, the magnetovolume shifts in both pure and Na-doped MnTe are linearly proportional to the LMOP, which gradually evolves into the AMOP below $T_{\mathrm{N}}$.

\section{Discussion}

The magnitude of a magnetostrictive response is determined by the balance between elastic energy costs and magnetic energy savings caused by changing the interatomic distances. In MnTe, the magnetic interactions are predominantly due to superexchange, with a smaller contribution from direct exchange~\cite{mu;prm19}. In either case, the interactions depend strongly on the relative positions of ions. The interactions between the first three nearest neighbors (see inset to Fig.~\ref{fig:MnTe_debyefit}) are $J_1 = -21.5$~K (along $c$), $J_2=0.67$~K (in the $ab$ plane), and $J_3=-2.87$~K (diagonally out of the plane)~\cite{szusz;prb06}. In superexchange systems where a single interaction along one crystalline axis dominates (as is the case with $J_1$ here), a simple energy minimization argument leads to the relation~\cite{zapf;prb08}
\begin{equation}
\label{eq:strainrelation}
    \epsilon_n = \frac{d_0}{VE}\frac{dJ}{dn}\langle\mathbf{S}_i\cdot \mathbf{S}_j \rangle,
\end{equation}
where $\epsilon_n$ is the strain along a given direction $\hat{n}$, $n$ is the coordinate along $\hat{n}$, $d_0$ is the equilibrium distance between neighboring magnetic ions in the absence of magnetism, $V$ is the unit cell volume, $E$ is the Young's modulus along $\hat{n}$, and $J$ is the magnetic interaction between the neighboring magnetic moments $\mathbf{S}_i$ and $\mathbf{S}_j$. Based on this equation, we can estimate the expected strain along the $c$ axis for MnTe. We know the values of $d_0$ and $V$ from the present work; the bulk modulus of MnTe has previously been reported as 47.3~GPa~\cite{paszk;appa97}, which we use for $E$; and we can estimate $dJ/dn$ to be $-5 \times 10^{-12}$~J/m based on tabulated superexchange interactions at various distances with similar cation-anion-cation angles as observed in MnTe~\cite{szusz;prb06}. Assuming fully ordered $S=5/2$ spins, the result is $\epsilon_c \sim 0.002$, which is remarkably close to the observed value of $\sim$0.003 at low temperature. Thus, we can attribute the exceptionally large magnetostriction along $c$ to a high density of large magnetic moments, a relatively low bulk modulus, and a strong spatial dependence of $J_1$. 

Often, a compressive strain along one direction will be partially counteracted by a tensile strain in other dimensions, minimizing the overall volume change. This has been observed, for example, in antiferromagnetic MnO~\cite{bloch;prb73} and FeTiO$_3$~\cite{chari;prb12} and in various ferromagnetic magnetocalorics~\cite{oey;apl20}. In MnTe, however, both $a$ and $c$ contract significantly as the magnetic correlations develop. We attribute this to the cooperative nature of the exchange interactions, such that no geometrical frustration is present to cause competing magnetoelastic tendencies. The favorable combination of material parameters leading to strong magnetostriction and the geometrical factors promoting the contraction of both $a$ and $c$ are therefore responsible for the exceptionally large magnetovolume effect in MnTe.

A more challenging issue is the clearly linear dependence of the lattice strain on the LMOP in MnTe, which stands in stark contrast to the expected
quadratic dependence of strain on the AMOP~\cite{pasca;jap82,chatt;jpcm10,chatt;jpcm11,carpe;jpcm12,chatt;jpcm12,oravo;jpcm13,miao;prb21}. This quadratic dependence originates from the time-reversal invariance of the free energy, which admits only even combined powers of any magnetic order parameters in the absence of an external magnetic field. We found the solution to this conundrum in the trilinear couplings of elastic strain with two distinct domains of the magnetic order parameter.

\begin{figure}
	\centerline{\includegraphics[width=75mm]{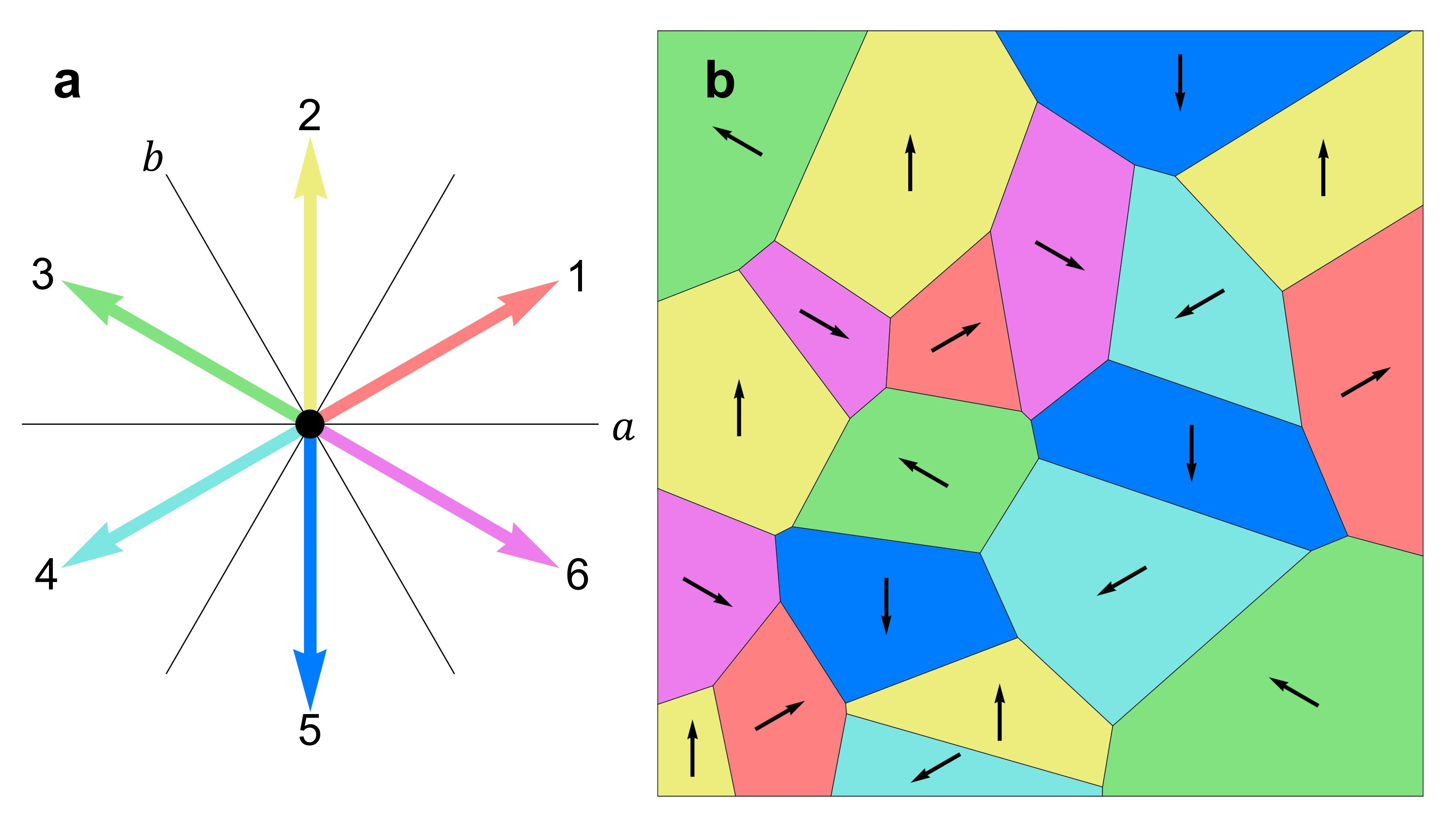}}
	\caption{\label{fig:domains} (a) The $z=0$ moment direction of each of the six magnetic domains of MnTe is indicated by a colored arrow. The directions marked \textit{a} and \textit{b} indicate the hexagonal lattice basis vectors. (b) Schematic representation of short-range-ordered magnetic correlations in the paramagnetic state of MnTe, where the $z=0$ moment direction indicated by the black arrow within a region of a given color matches the magnetic domain with an arrow of the same color in (a). The magnetic-domain pattern is intended to represent dynamic rather than static short-range magnetic correlations. }		
\end{figure}

The AFM structure of MnTe consists of alternating ferromagnetic layers of in-plane moments oriented between the in-plane lattice basis vectors, so that moments in the $z=0$ and $z=0.5$ layers have opposite directions~\cite{krieg;nc16,mosel;prm22,baral;matter22}. The magnetic order parameter belongs to irreducible representation (irrep) and order parameter direction (OPD) $\Gamma_5^+(0,a)$ of non-magnetic parent space-group $P6_3/mmc$ (\#194), which yields magnetic space group $Cm'c'm$ (BNS~63.462). This magnetic order parameter has six distinct but symmetry-equivalent domains characterized by the moment direction in the $z=0$ layer, as shown in Fig.~\ref{fig:domains}(a).
In this context, short-range magnetic order can be viewed as a dynamically fluctuating spatial distribution of small regions, each possessing one of the domains of the order parameter.  Below $T_{\mathrm{N}}$, one of these six domains emerges as the long-range order parameter, while the other domains gradually cease to be represented. We define $m_S$ to be the short-range magnetic order parameter corresponding to any one of the six magnetic domains and $m_L$ to be the emerging long-range magnetic order parameter, which we will arbitrarily associate with domain \#1 in Fig.~\ref{fig:domains}(a). Referring to the insets in Fig.~\ref{fig:MnTe_debyefit}, we associate $m_S$ with the approximately temperature-independent correlations of magnitude $\sim 1$~$\mu_B$ and correlation length $\sim 3$~\AA\ observed above 375~K, while the onset of $m_L$ is identified by the rapid rise in correlation length below 375~K, leading to genuine long-range order below $T_{\mathrm{N}}$. 

As the length scale and lifetime of $m_L$ increase upon cooling toward $T_{\mathrm{N}}$, the superposition of $m_L$ and $m_S$ becomes important. Each combination of a strain variable $\epsilon$ with a domain $S$ of the short-range magnetic order parameter contributes a symmetry-allowed trilinear coupling term of the form $\epsilon m_L m_S$ to the free energy, where the coefficient of the term depends on the angle between the orientations of their $z=0$ moment directions from Fig.~\ref{fig:domains}(a). Minimizing the free energy, which also includes an $\epsilon^2$ term, with respect to strain then yields $\epsilon \sim m_L m_S$. The key observation is that the LMOP ($m_S$) is already large and relatively temperature insensitive when the magnetostriction begins around 375~K, so that it can be treated as a \textit{pre-existing} order parameter. This provides a symmetry-allowed mechanism through which $\epsilon$ can track linearly with $m_L$, thus leading to the observed linear magnetostrictive response. This reasoning also applies to the magnetic structure of Na-doped MnTe, as demonstrated in the SI.

To our knowledge, linear spontaneous magnetostriction has not been reported previously, making MnTe unique both for the exceptionally large magnitude and the linear nature of the spontaneous magnetostriction. However, we expect that other materials with a correlated paramagnetic state and suitable domain couplings may also exhibit such behavior, making this perhaps a more widespread mechanism for spontaneous magnetostriction. This work also motivates further study of the oft-overlooked role of short-range magnetic order in magnetostructural effects and the coupling of domains of short-range order parameters in general.


\textbf{Acknowledgements}

We thank Chris Howard for helpful discussions regarding magnetostructural coupling. B.A.F. and R.B. were supported by the U.S. Department of Energy, Office of Science, Basic Energy Science through Award No. DE-SC0021134. This research used beamline 28-ID-1 of the National Synchrotron Light Source II, a U.S. Department of Energy (DOE) Office of Science User Facility operated for the DOE Office of Science by Brookhaven National Laboratory under Contract No. DE-SC0012704. 

\textbf{Conflict of Interest}

The authors declare no conflict of interest.


\begin{thebibliography}{53}%
	\makeatletter
	\providecommand \@ifxundefined [1]{%
		\@ifx{#1\undefined}
	}%
	\providecommand \@ifnum [1]{%
		\ifnum #1\expandafter \@firstoftwo
		\else \expandafter \@secondoftwo
		\fi
	}%
	\providecommand \@ifx [1]{%
		\ifx #1\expandafter \@firstoftwo
		\else \expandafter \@secondoftwo
		\fi
	}%
	\providecommand \natexlab [1]{#1}%
	\providecommand \enquote  [1]{``#1''}%
	\providecommand \bibnamefont  [1]{#1}%
	\providecommand \bibfnamefont [1]{#1}%
	\providecommand \citenamefont [1]{#1}%
	\providecommand \href@noop [0]{\@secondoftwo}%
	\providecommand \href [0]{\begingroup \@sanitize@url \@href}%
	\providecommand \@href[1]{\@@startlink{#1}\@@href}%
	\providecommand \@@href[1]{\endgroup#1\@@endlink}%
	\providecommand \@sanitize@url [0]{\catcode `\\12\catcode `\$12\catcode
		`\&12\catcode `\#12\catcode `\^12\catcode `\_12\catcode `\%12\relax}%
	\providecommand \@@startlink[1]{}%
	\providecommand \@@endlink[0]{}%
	\providecommand \url  [0]{\begingroup\@sanitize@url \@url }%
	\providecommand \@url [1]{\endgroup\@href {#1}{\urlprefix }}%
	\providecommand \urlprefix  [0]{URL }%
	\providecommand \Eprint [0]{\href }%
	\providecommand \doibase [0]{https://doi.org/}%
	\providecommand \selectlanguage [0]{\@gobble}%
	\providecommand \bibinfo  [0]{\@secondoftwo}%
	\providecommand \bibfield  [0]{\@secondoftwo}%
	\providecommand \translation [1]{[#1]}%
	\providecommand \BibitemOpen [0]{}%
	\providecommand \bibitemStop [0]{}%
	\providecommand \bibitemNoStop [0]{.\EOS\space}%
	\providecommand \EOS [0]{\spacefactor3000\relax}%
	\providecommand \BibitemShut  [1]{\csname bibitem#1\endcsname}%
	\let\auto@bib@innerbib\@empty
	\bibitem [{\citenamefont {Johnston}\ and\ \citenamefont
		{Sestrich}(1961)}]{johns;jinc61}%
	\BibitemOpen
	\bibfield  {author} {\bibinfo {author} {\bibfnamefont {W.~D.}\ \bibnamefont
			{Johnston}}\ and\ \bibinfo {author} {\bibfnamefont {D.~E.}\ \bibnamefont
			{Sestrich}},\ }\bibfield  {title} {\bibinfo {title} {{The MnTe--GeTe phase
				diagram}},\ }\href
	{https://doi.org/https://doi.org/10.1016/0022-1902(61)80111-5} {\bibfield
		{journal} {\bibinfo  {journal} {J. Inorg. Nucl. Chem.}\ }\textbf {\bibinfo
			{volume} {19}},\ \bibinfo {pages} {229} (\bibinfo {year} {1961})}\BibitemShut
	{NoStop}%
	\bibitem [{\citenamefont {Kriegner}\ \emph {et~al.}(2016)\citenamefont
		{Kriegner}, \citenamefont {V\'yborn\'y}, \citenamefont {Olejn\'ik},
		\citenamefont {Reichlov\'a}, \citenamefont {Nov\'ak}, \citenamefont {Marti},
		\citenamefont {Gazquez}, \citenamefont {Saidl}, \citenamefont {N\v{e}mec},
		\citenamefont {Volobuev}, \citenamefont {Springholz}, \citenamefont
		{Hol\'y},\ and\ \citenamefont {Jungwirth}}]{krieg;nc16}%
	\BibitemOpen
	\bibfield  {author} {\bibinfo {author} {\bibfnamefont {D.}~\bibnamefont
			{Kriegner}}, \bibinfo {author} {\bibfnamefont {K.}~\bibnamefont
			{V\'yborn\'y}}, \bibinfo {author} {\bibfnamefont {K.}~\bibnamefont
			{Olejn\'ik}}, \bibinfo {author} {\bibfnamefont {H.}~\bibnamefont
			{Reichlov\'a}}, \bibinfo {author} {\bibfnamefont {V.}~\bibnamefont
			{Nov\'ak}}, \bibinfo {author} {\bibfnamefont {X.}~\bibnamefont {Marti}},
		\bibinfo {author} {\bibfnamefont {J.}~\bibnamefont {Gazquez}}, \bibinfo
		{author} {\bibfnamefont {V.}~\bibnamefont {Saidl}}, \bibinfo {author}
		{\bibfnamefont {P.}~\bibnamefont {N\v{e}mec}}, \bibinfo {author}
		{\bibfnamefont {V.~V.}\ \bibnamefont {Volobuev}}, \bibinfo {author}
		{\bibfnamefont {G.}~\bibnamefont {Springholz}}, \bibinfo {author}
		{\bibfnamefont {V.}~\bibnamefont {Hol\'y}},\ and\ \bibinfo {author}
		{\bibfnamefont {T.}~\bibnamefont {Jungwirth}},\ }\bibfield  {title} {\bibinfo
		{title} {{Multiple-stable anisotropic magnetoresistance memory in
				antiferromagnetic MnTe}},\ }\href@noop {} {\bibfield  {journal} {\bibinfo
			{journal} {Nat. Commun.}\ }\textbf {\bibinfo {volume} {7}},\ \bibinfo {pages}
		{11623} (\bibinfo {year} {2016})}\BibitemShut {NoStop}%
	\bibitem [{\citenamefont {Yin}\ \emph {et~al.}(2019)\citenamefont {Yin},
		\citenamefont {Yu}, \citenamefont {Liu}, \citenamefont {Lake}, \citenamefont
		{Zang},\ and\ \citenamefont {Wang}}]{yin;prl19}%
	\BibitemOpen
	\bibfield  {author} {\bibinfo {author} {\bibfnamefont {G.}~\bibnamefont
			{Yin}}, \bibinfo {author} {\bibfnamefont {J.-X.}\ \bibnamefont {Yu}},
		\bibinfo {author} {\bibfnamefont {Y.}~\bibnamefont {Liu}}, \bibinfo {author}
		{\bibfnamefont {R.~K.}\ \bibnamefont {Lake}}, \bibinfo {author}
		{\bibfnamefont {J.}~\bibnamefont {Zang}},\ and\ \bibinfo {author}
		{\bibfnamefont {K.~L.}\ \bibnamefont {Wang}},\ }\bibfield  {title} {\bibinfo
		{title} {{Planar Hall Effect in Antiferromagnetic MnTe Thin Films}},\ }\href
	{https://doi.org/10.1103/PhysRevLett.122.106602} {\bibfield  {journal}
		{\bibinfo  {journal} {Phys. Rev. Lett.}\ }\textbf {\bibinfo {volume} {122}},\
		\bibinfo {pages} {106602} (\bibinfo {year} {2019})}\BibitemShut {NoStop}%
	\bibitem [{\citenamefont {Bossini}\ \emph {et~al.}(2021)\citenamefont
		{Bossini}, \citenamefont {Dal~Conte}, \citenamefont {Terschanski},
		\citenamefont {Springholz}, \citenamefont {Bonanni}, \citenamefont
		{Deltenre}, \citenamefont {Anders}, \citenamefont {Uhrig}, \citenamefont
		{Cerullo},\ and\ \citenamefont {Cinchetti}}]{bossi;prb21}%
	\BibitemOpen
	\bibfield  {author} {\bibinfo {author} {\bibfnamefont {D.}~\bibnamefont
			{Bossini}}, \bibinfo {author} {\bibfnamefont {S.}~\bibnamefont {Dal~Conte}},
		\bibinfo {author} {\bibfnamefont {M.}~\bibnamefont {Terschanski}}, \bibinfo
		{author} {\bibfnamefont {G.}~\bibnamefont {Springholz}}, \bibinfo {author}
		{\bibfnamefont {A.}~\bibnamefont {Bonanni}}, \bibinfo {author} {\bibfnamefont
			{K.}~\bibnamefont {Deltenre}}, \bibinfo {author} {\bibfnamefont
			{F.}~\bibnamefont {Anders}}, \bibinfo {author} {\bibfnamefont {G.~S.}\
			\bibnamefont {Uhrig}}, \bibinfo {author} {\bibfnamefont {G.}~\bibnamefont
			{Cerullo}},\ and\ \bibinfo {author} {\bibfnamefont {M.}~\bibnamefont
			{Cinchetti}},\ }\bibfield  {title} {\bibinfo {title} {Femtosecond phononic
			coupling to both spins and charges in a room-temperature antiferromagnetic
			semiconductor},\ }\href {https://doi.org/10.1103/PhysRevB.104.224424}
	{\bibfield  {journal} {\bibinfo  {journal} {Phys. Rev. B}\ }\textbf {\bibinfo
			{volume} {104}},\ \bibinfo {pages} {224424} (\bibinfo {year}
		{2021})}\BibitemShut {NoStop}%
	\bibitem [{\citenamefont {Xu}\ \emph {et~al.}(2017)\citenamefont {Xu},
		\citenamefont {Li}, \citenamefont {Wang}, \citenamefont {Li}, \citenamefont
		{Chen}, \citenamefont {Lin}, \citenamefont {Chen},\ and\ \citenamefont
		{Pei}}]{xu;jmchema17}%
	\BibitemOpen
	\bibfield  {author} {\bibinfo {author} {\bibfnamefont {Y.}~\bibnamefont
			{Xu}}, \bibinfo {author} {\bibfnamefont {W.}~\bibnamefont {Li}}, \bibinfo
		{author} {\bibfnamefont {C.}~\bibnamefont {Wang}}, \bibinfo {author}
		{\bibfnamefont {J.}~\bibnamefont {Li}}, \bibinfo {author} {\bibfnamefont
			{Z.}~\bibnamefont {Chen}}, \bibinfo {author} {\bibfnamefont {S.}~\bibnamefont
			{Lin}}, \bibinfo {author} {\bibfnamefont {Y.}~\bibnamefont {Chen}},\ and\
		\bibinfo {author} {\bibfnamefont {Y.}~\bibnamefont {Pei}},\ }\bibfield
	{title} {\bibinfo {title} {{Performance optimization and single parabolic
				band behavior of thermoelectric MnTe}},\ }\href
	{https://doi.org/10.1039/C7TA04842D} {\bibfield  {journal} {\bibinfo
			{journal} {J. Mater. Chem. A}\ }\textbf {\bibinfo {volume} {5}},\ \bibinfo
		{pages} {19143} (\bibinfo {year} {2017})}\BibitemShut {NoStop}%
	\bibitem [{\citenamefont {Ren}\ \emph {et~al.}(2017)\citenamefont {Ren},
		\citenamefont {Yang}, \citenamefont {Jiang}, \citenamefont {Zhang},
		\citenamefont {Zhou}, \citenamefont {Li}, \citenamefont {Xin},\ and\
		\citenamefont {He}}]{ren;jmchemc17}%
	\BibitemOpen
	\bibfield  {author} {\bibinfo {author} {\bibfnamefont {Y.}~\bibnamefont
			{Ren}}, \bibinfo {author} {\bibfnamefont {J.}~\bibnamefont {Yang}}, \bibinfo
		{author} {\bibfnamefont {Q.}~\bibnamefont {Jiang}}, \bibinfo {author}
		{\bibfnamefont {D.}~\bibnamefont {Zhang}}, \bibinfo {author} {\bibfnamefont
			{Z.}~\bibnamefont {Zhou}}, \bibinfo {author} {\bibfnamefont {X.}~\bibnamefont
			{Li}}, \bibinfo {author} {\bibfnamefont {J.}~\bibnamefont {Xin}},\ and\
		\bibinfo {author} {\bibfnamefont {X.}~\bibnamefont {He}},\ }\bibfield
	{title} {\bibinfo {title} {{Synergistic effect by Na doping and S
				substitution for high thermoelectric performance of p-type MnTe}},\ }\href
	{https://doi.org/10.1039/c7tc01480e} {\bibfield  {journal} {\bibinfo
			{journal} {J. Mater. Chem. C}\ }\textbf {\bibinfo {volume} {5}},\ \bibinfo
		{pages} {5076} (\bibinfo {year} {2017})}\BibitemShut {NoStop}%
	\bibitem [{\citenamefont {Dong}\ \emph {et~al.}(2018)\citenamefont {Dong},
		\citenamefont {Wu}, \citenamefont {Pei}, \citenamefont {Sun}, \citenamefont
		{Pan}, \citenamefont {Zhang}, \citenamefont {Tang},\ and\ \citenamefont
		{Li}}]{dong;jmchemc18}%
	\BibitemOpen
	\bibfield  {author} {\bibinfo {author} {\bibfnamefont {J.}~\bibnamefont
			{Dong}}, \bibinfo {author} {\bibfnamefont {C.-F.}\ \bibnamefont {Wu}},
		\bibinfo {author} {\bibfnamefont {J.}~\bibnamefont {Pei}}, \bibinfo {author}
		{\bibfnamefont {F.-H.}\ \bibnamefont {Sun}}, \bibinfo {author} {\bibfnamefont
			{Y.}~\bibnamefont {Pan}}, \bibinfo {author} {\bibfnamefont {B.-P.}\
			\bibnamefont {Zhang}}, \bibinfo {author} {\bibfnamefont {H.}~\bibnamefont
			{Tang}},\ and\ \bibinfo {author} {\bibfnamefont {J.-F.}\ \bibnamefont {Li}},\
	}\bibfield  {title} {\bibinfo {title} {{Lead-free MnTe mid-temperature
				thermoelectric materials: facile synthesis, p-type doping and transport
				properties}},\ }\href {https://doi.org/10.1039/C8TC00904J} {\bibfield
		{journal} {\bibinfo  {journal} {J. Mater. Chem. C}\ }\textbf {\bibinfo
			{volume} {6}},\ \bibinfo {pages} {4265} (\bibinfo {year} {2018})}\BibitemShut
	{NoStop}%
	\bibitem [{\citenamefont {Deng}\ \emph {et~al.}(2021)\citenamefont {Deng},
		\citenamefont {Lou}, \citenamefont {Lu}, \citenamefont {Zhang}, \citenamefont
		{Li}, \citenamefont {Li}, \citenamefont {Zhang}, \citenamefont {Zhang},
		\citenamefont {Chen}, \citenamefont {Zhang}, \citenamefont {Zhang},\ and\
		\citenamefont {Tang}}]{deng;nanoe21}%
	\BibitemOpen
	\bibfield  {author} {\bibinfo {author} {\bibfnamefont {H.}~\bibnamefont
			{Deng}}, \bibinfo {author} {\bibfnamefont {X.}~\bibnamefont {Lou}}, \bibinfo
		{author} {\bibfnamefont {W.}~\bibnamefont {Lu}}, \bibinfo {author}
		{\bibfnamefont {J.}~\bibnamefont {Zhang}}, \bibinfo {author} {\bibfnamefont
			{D.}~\bibnamefont {Li}}, \bibinfo {author} {\bibfnamefont {S.}~\bibnamefont
			{Li}}, \bibinfo {author} {\bibfnamefont {Q.}~\bibnamefont {Zhang}}, \bibinfo
		{author} {\bibfnamefont {X.}~\bibnamefont {Zhang}}, \bibinfo {author}
		{\bibfnamefont {X.}~\bibnamefont {Chen}}, \bibinfo {author} {\bibfnamefont
			{D.}~\bibnamefont {Zhang}}, \bibinfo {author} {\bibfnamefont
			{Y.}~\bibnamefont {Zhang}},\ and\ \bibinfo {author} {\bibfnamefont
			{G.}~\bibnamefont {Tang}},\ }\bibfield  {title} {\bibinfo {title}
		{{High-performance eco-friendly MnTe thermoelectrics through introducing SnTe
				nanocrystals and manipulating band structure}},\ }\href
	{https://doi.org/https://doi.org/10.1016/j.nanoen.2020.105649} {\bibfield
		{journal} {\bibinfo  {journal} {Nano Energy}\ }\textbf {\bibinfo {volume}
			{81}},\ \bibinfo {pages} {105649} (\bibinfo {year} {2021})}\BibitemShut
	{NoStop}%
	\bibitem [{\citenamefont {Xiong}\ \emph {et~al.}(2023)\citenamefont {Xiong},
		\citenamefont {Wang}, \citenamefont {Zhang}, \citenamefont {Wang},
		\citenamefont {Yin}, \citenamefont {Gong}, \citenamefont {Zhang},
		\citenamefont {Li}, \citenamefont {Liu}, \citenamefont {Wang}, \citenamefont
		{Zhang},\ and\ \citenamefont {Tang}}]{xiong;sma23}%
	\BibitemOpen
	\bibfield  {author} {\bibinfo {author} {\bibfnamefont {W.}~\bibnamefont
			{Xiong}}, \bibinfo {author} {\bibfnamefont {Z.}~\bibnamefont {Wang}},
		\bibinfo {author} {\bibfnamefont {X.}~\bibnamefont {Zhang}}, \bibinfo
		{author} {\bibfnamefont {C.}~\bibnamefont {Wang}}, \bibinfo {author}
		{\bibfnamefont {L.}~\bibnamefont {Yin}}, \bibinfo {author} {\bibfnamefont
			{Y.}~\bibnamefont {Gong}}, \bibinfo {author} {\bibfnamefont {Q.}~\bibnamefont
			{Zhang}}, \bibinfo {author} {\bibfnamefont {S.}~\bibnamefont {Li}}, \bibinfo
		{author} {\bibfnamefont {Q.}~\bibnamefont {Liu}}, \bibinfo {author}
		{\bibfnamefont {P.}~\bibnamefont {Wang}}, \bibinfo {author} {\bibfnamefont
			{Y.}~\bibnamefont {Zhang}},\ and\ \bibinfo {author} {\bibfnamefont
			{G.}~\bibnamefont {Tang}},\ }\bibfield  {title} {\bibinfo {title} {{Lattice
				Distortions and Multiple Valence Band Convergence Contributing to High
				Thermoelectric Performance in MnTe}},\ }\href
	{https://doi.org/https://doi.org/10.1002/smll.202206058} {\bibfield
		{journal} {\bibinfo  {journal} {Small}\ }\textbf {\bibinfo {volume} {19}},\
		\bibinfo {pages} {2206058} (\bibinfo {year} {2023})}\BibitemShut {NoStop}%
	\bibitem [{\citenamefont {Zulkifal}\ \emph {et~al.}(2023)\citenamefont
		{Zulkifal}, \citenamefont {Wang}, \citenamefont {Zhang}, \citenamefont
		{Siddique}, \citenamefont {Yu}, \citenamefont {Wang}, \citenamefont {Gong},
		\citenamefont {Li}, \citenamefont {Li}, \citenamefont {Zhang}, \citenamefont
		{Wang},\ and\ \citenamefont {Tang}}]{zulki;advs23}%
	\BibitemOpen
	\bibfield  {author} {\bibinfo {author} {\bibfnamefont {S.}~\bibnamefont
			{Zulkifal}}, \bibinfo {author} {\bibfnamefont {Z.}~\bibnamefont {Wang}},
		\bibinfo {author} {\bibfnamefont {X.}~\bibnamefont {Zhang}}, \bibinfo
		{author} {\bibfnamefont {S.}~\bibnamefont {Siddique}}, \bibinfo {author}
		{\bibfnamefont {Y.}~\bibnamefont {Yu}}, \bibinfo {author} {\bibfnamefont
			{C.}~\bibnamefont {Wang}}, \bibinfo {author} {\bibfnamefont {Y.}~\bibnamefont
			{Gong}}, \bibinfo {author} {\bibfnamefont {S.}~\bibnamefont {Li}}, \bibinfo
		{author} {\bibfnamefont {D.}~\bibnamefont {Li}}, \bibinfo {author}
		{\bibfnamefont {Y.}~\bibnamefont {Zhang}}, \bibinfo {author} {\bibfnamefont
			{P.}~\bibnamefont {Wang}},\ and\ \bibinfo {author} {\bibfnamefont
			{G.}~\bibnamefont {Tang}},\ }\bibfield  {title} {\bibinfo {title} {Multiple
			valence bands convergence and localized lattice engineering lead to superhigh
			thermoelectric figure of merit in mnte},\ }\href
	{https://doi.org/https://doi.org/10.1002/advs.202206342} {\bibfield
		{journal} {\bibinfo  {journal} {Adv. Sci.}\ ,\ \bibinfo {pages} {2206342}}
		(\bibinfo {year} {2023})}\BibitemShut {NoStop}%
	\bibitem [{\citenamefont {Zheng}\ \emph {et~al.}(2019)\citenamefont {Zheng},
		\citenamefont {Lu}, \citenamefont {Polash}, \citenamefont
		{Rasoulianboroujeni}, \citenamefont {Liu}, \citenamefont {Manley},
		\citenamefont {Deng}, \citenamefont {Sun}, \citenamefont {Chen},
		\citenamefont {Hermann}, \citenamefont {Vashaee}, \citenamefont {Heremans},\
		and\ \citenamefont {Zhao}}]{zheng;sadv19}%
	\BibitemOpen
	\bibfield  {author} {\bibinfo {author} {\bibfnamefont {Y.}~\bibnamefont
			{Zheng}}, \bibinfo {author} {\bibfnamefont {T.}~\bibnamefont {Lu}}, \bibinfo
		{author} {\bibfnamefont {M.~M.~H.}\ \bibnamefont {Polash}}, \bibinfo {author}
		{\bibfnamefont {M.}~\bibnamefont {Rasoulianboroujeni}}, \bibinfo {author}
		{\bibfnamefont {N.}~\bibnamefont {Liu}}, \bibinfo {author} {\bibfnamefont
			{M.~E.}\ \bibnamefont {Manley}}, \bibinfo {author} {\bibfnamefont
			{Y.}~\bibnamefont {Deng}}, \bibinfo {author} {\bibfnamefont {P.~J.}\
			\bibnamefont {Sun}}, \bibinfo {author} {\bibfnamefont {X.~L.}\ \bibnamefont
			{Chen}}, \bibinfo {author} {\bibfnamefont {R.~P.}\ \bibnamefont {Hermann}},
		\bibinfo {author} {\bibfnamefont {D.}~\bibnamefont {Vashaee}}, \bibinfo
		{author} {\bibfnamefont {J.~P.}\ \bibnamefont {Heremans}},\ and\ \bibinfo
		{author} {\bibfnamefont {H.}~\bibnamefont {Zhao}},\ }\bibfield  {title}
	{\bibinfo {title} {{Paramagnon drag in high thermoelectric figure of merit
				Li-doped MnTe}},\ }\href {https://doi.org/10.1126/sciadv.aat9461} {\bibfield
		{journal} {\bibinfo  {journal} {Sci. Adv.}\ }\textbf {\bibinfo {volume}
			{5}},\ \bibinfo {pages} {eaat9461} (\bibinfo {year} {2019})}\BibitemShut
	{NoStop}%
	\bibitem [{\citenamefont {Polash}\ \emph {et~al.}(2021)\citenamefont {Polash},
		\citenamefont {Moseley}, \citenamefont {Zhang}, \citenamefont {Hermann},\
		and\ \citenamefont {Vashaee}}]{polas;crps21}%
	\BibitemOpen
	\bibfield  {author} {\bibinfo {author} {\bibfnamefont {M.~M.~H.}\
			\bibnamefont {Polash}}, \bibinfo {author} {\bibfnamefont {D.}~\bibnamefont
			{Moseley}}, \bibinfo {author} {\bibfnamefont {J.}~\bibnamefont {Zhang}},
		\bibinfo {author} {\bibfnamefont {R.~P.}\ \bibnamefont {Hermann}},\ and\
		\bibinfo {author} {\bibfnamefont {D.}~\bibnamefont {Vashaee}},\ }\bibfield
	{title} {\bibinfo {title} {{Understanding and design of spin-driven
				thermoelectrics}},\ }\href
	{https://doi.org/https://doi.org/10.1016/j.xcrp.2021.100614} {\bibfield
		{journal} {\bibinfo  {journal} {Cell Rep. Phys. Sci.}\ }\textbf {\bibinfo
			{volume} {2}},\ \bibinfo {pages} {100614} (\bibinfo {year}
		{2021})}\BibitemShut {NoStop}%
	\bibitem [{\citenamefont {Baral}\ \emph {et~al.}(2022)\citenamefont {Baral},
		\citenamefont {Christensen}, \citenamefont {Hamilton}, \citenamefont {Ye},
		\citenamefont {Chesnel}, \citenamefont {Sparks}, \citenamefont {Ward},
		\citenamefont {Yan}, \citenamefont {McGuire}, \citenamefont {Manley},
		\citenamefont {Staunton}, \citenamefont {Hermann},\ and\ \citenamefont
		{Frandsen}}]{baral;matter22}%
	\BibitemOpen
	\bibfield  {author} {\bibinfo {author} {\bibfnamefont {R.}~\bibnamefont
			{Baral}}, \bibinfo {author} {\bibfnamefont {J.}~\bibnamefont {Christensen}},
		\bibinfo {author} {\bibfnamefont {P.}~\bibnamefont {Hamilton}}, \bibinfo
		{author} {\bibfnamefont {F.}~\bibnamefont {Ye}}, \bibinfo {author}
		{\bibfnamefont {K.}~\bibnamefont {Chesnel}}, \bibinfo {author} {\bibfnamefont
			{T.~D.}\ \bibnamefont {Sparks}}, \bibinfo {author} {\bibfnamefont
			{R.}~\bibnamefont {Ward}}, \bibinfo {author} {\bibfnamefont {J.}~\bibnamefont
			{Yan}}, \bibinfo {author} {\bibfnamefont {M.~A.}\ \bibnamefont {McGuire}},
		\bibinfo {author} {\bibfnamefont {M.~E.}\ \bibnamefont {Manley}}, \bibinfo
		{author} {\bibfnamefont {J.~B.}\ \bibnamefont {Staunton}}, \bibinfo {author}
		{\bibfnamefont {R.~P.}\ \bibnamefont {Hermann}},\ and\ \bibinfo {author}
		{\bibfnamefont {B.~A.}\ \bibnamefont {Frandsen}},\ }\bibfield  {title}
	{\bibinfo {title} {Real-space visualization of short-range antiferromagnetic
			correlations in a magnetically enhanced thermoelectric},\ }\href
	{https://doi.org/10.1016/j.matt.2022.03.011} {\bibfield  {journal} {\bibinfo
			{journal} {Matter}\ }\textbf {\bibinfo {volume} {5}},\ \bibinfo {pages}
		{1853} (\bibinfo {year} {2022})}\BibitemShut {NoStop}%
	\bibitem [{\citenamefont {Deng}\ \emph {et~al.}(2020)\citenamefont {Deng},
		\citenamefont {Yu}, \citenamefont {Shi}, \citenamefont {Guo}, \citenamefont
		{Xu}, \citenamefont {Wang}, \citenamefont {Chen},\ and\ \citenamefont
		{Zhang}}]{deng;s20}%
	\BibitemOpen
	\bibfield  {author} {\bibinfo {author} {\bibfnamefont {Y.}~\bibnamefont
			{Deng}}, \bibinfo {author} {\bibfnamefont {Y.}~\bibnamefont {Yu}}, \bibinfo
		{author} {\bibfnamefont {M.~Z.}\ \bibnamefont {Shi}}, \bibinfo {author}
		{\bibfnamefont {Z.}~\bibnamefont {Guo}}, \bibinfo {author} {\bibfnamefont
			{Z.}~\bibnamefont {Xu}}, \bibinfo {author} {\bibfnamefont {J.}~\bibnamefont
			{Wang}}, \bibinfo {author} {\bibfnamefont {X.~H.}\ \bibnamefont {Chen}},\
		and\ \bibinfo {author} {\bibfnamefont {Y.}~\bibnamefont {Zhang}},\ }\bibfield
	{title} {\bibinfo {title} {{Quantum anomalous Hall effect in intrinsic
				magnetic topological insulator MnBi$_2$Te$_4$}},\ }\href
	{https://doi.org/10.1126/science.aax8156} {\bibfield  {journal} {\bibinfo
			{journal} {Science}\ }\textbf {\bibinfo {volume} {367}},\ \bibinfo {pages}
		{895} (\bibinfo {year} {2020})}\BibitemShut {NoStop}%
	\bibitem [{\citenamefont {Allen}\ \emph {et~al.}(1977)\citenamefont {Allen},
		\citenamefont {Lucovsky},\ and\ \citenamefont {Mikkelsen~Jr.}}]{allen;ssc77}%
	\BibitemOpen
	\bibfield  {author} {\bibinfo {author} {\bibfnamefont {J.~W.}\ \bibnamefont
			{Allen}}, \bibinfo {author} {\bibfnamefont {G.}~\bibnamefont {Lucovsky}},\
		and\ \bibinfo {author} {\bibfnamefont {J.~C.}\ \bibnamefont
			{Mikkelsen~Jr.}},\ }\bibfield  {title} {\bibinfo {title} {{Optical Properties
				and Electronic Structure of Crossroads Material MnTe}},\ }\href@noop {}
	{\bibfield  {journal} {\bibinfo  {journal} {Solid State Commun.}\ }\textbf
		{\bibinfo {volume} {24}},\ \bibinfo {pages} {367} (\bibinfo {year}
		{1977})}\BibitemShut {NoStop}%
	\bibitem [{\citenamefont {Youn}\ \emph {et~al.}(2004)\citenamefont {Youn},
		\citenamefont {Min},\ and\ \citenamefont {Freeman}}]{youn;pssb04}%
	\BibitemOpen
	\bibfield  {author} {\bibinfo {author} {\bibfnamefont {S.~J.}\ \bibnamefont
			{Youn}}, \bibinfo {author} {\bibfnamefont {B.~I.}\ \bibnamefont {Min}},\ and\
		\bibinfo {author} {\bibfnamefont {A.~J.}\ \bibnamefont {Freeman}},\
	}\bibfield  {title} {\bibinfo {title} {{Crossroads electronic structure of
				MnS, MnSe, and MnTe}},\ }\href
	{https://doi.org/https://doi.org/10.1002/pssb.200304538} {\bibfield
		{journal} {\bibinfo  {journal} {Phys. Status Solidi B}\ }\textbf {\bibinfo
			{volume} {241}},\ \bibinfo {pages} {1411} (\bibinfo {year}
		{2004})}\BibitemShut {NoStop}%
	\bibitem [{\citenamefont {Du~Tr\'{e}molet~de
			Lacheisserie}(1993)}]{dutre;b;mtaaom93}%
	\BibitemOpen
	\bibfield  {author} {\bibinfo {author} {\bibfnamefont {E.}~\bibnamefont
			{Du~Tr\'{e}molet~de Lacheisserie}},\ }\href@noop {} {\emph {\bibinfo {title}
			{Magnetostriction: theory and applications of magnetoelasticity}}}\ (\bibinfo
	{publisher} {CRC Press},\ \bibinfo {address} {Boca Raton},\ \bibinfo {year}
	{1993})\BibitemShut {NoStop}%
	\bibitem [{\citenamefont {Hristoforou}\ and\ \citenamefont
		{Ktena}(2007)}]{hrist;jmmm07}%
	\BibitemOpen
	\bibfield  {author} {\bibinfo {author} {\bibfnamefont {E.}~\bibnamefont
			{Hristoforou}}\ and\ \bibinfo {author} {\bibfnamefont {A.}~\bibnamefont
			{Ktena}},\ }\bibfield  {title} {\bibinfo {title} {Magnetostriction and
			magnetostrictive materials for sensing applications},\ }\href
	{https://doi.org/https://doi.org/10.1016/j.jmmm.2007.03.025} {\bibfield
		{journal} {\bibinfo  {journal} {J. Magn. Magn. Mater.}\ }\textbf {\bibinfo
			{volume} {316}},\ \bibinfo {pages} {372} (\bibinfo {year} {2007})},\ \bibinfo
	{note} {proceedings of the Joint European Magnetic Symposia}\BibitemShut
	{NoStop}%
	\bibitem [{\citenamefont {Gao}\ \emph {et~al.}(2022)\citenamefont {Gao},
		\citenamefont {Zeng}, \citenamefont {Peng},\ and\ \citenamefont
		{Shuai}}]{gao;bam22}%
	\BibitemOpen
	\bibfield  {author} {\bibinfo {author} {\bibfnamefont {C.}~\bibnamefont
			{Gao}}, \bibinfo {author} {\bibfnamefont {Z.}~\bibnamefont {Zeng}}, \bibinfo
		{author} {\bibfnamefont {S.}~\bibnamefont {Peng}},\ and\ \bibinfo {author}
		{\bibfnamefont {C.}~\bibnamefont {Shuai}},\ }\bibfield  {title} {\bibinfo
		{title} {{Magnetostrictive alloys: Promising materials for biomedical
				applications}},\ }\href
	{https://doi.org/https://doi.org/10.1016/j.bioactmat.2021.06.025} {\bibfield
		{journal} {\bibinfo  {journal} {Bioactive Materials}\ }\textbf {\bibinfo
			{volume} {8}},\ \bibinfo {pages} {177} (\bibinfo {year} {2022})}\BibitemShut
	{NoStop}%
	\bibitem [{\citenamefont {Angus}\ and\ \citenamefont
		{Neppiras}(1969)}]{angus;u69}%
	\BibitemOpen
	\bibfield  {author} {\bibinfo {author} {\bibfnamefont {H.}~\bibnamefont
			{Angus}}\ and\ \bibinfo {author} {\bibfnamefont {E.}~\bibnamefont
			{Neppiras}},\ }\bibfield  {title} {\bibinfo {title} {Nickel-based
			magnetostrictive alloys for electromechanical transducers},\ }\href
	{https://doi.org/https://doi.org/10.1016/0041-624X(69)90664-7} {\bibfield
		{journal} {\bibinfo  {journal} {Ultrasonics}\ }\textbf {\bibinfo {volume}
			{7}},\ \bibinfo {pages} {182} (\bibinfo {year} {1969})}\BibitemShut {NoStop}%
	\bibitem [{\citenamefont {Jiles}(1994)}]{jile;jpd94}%
	\BibitemOpen
	\bibfield  {author} {\bibinfo {author} {\bibfnamefont {D.}~\bibnamefont
			{Jiles}},\ }\bibfield  {title} {\bibinfo {title} {The development of highly
			magnetostrictive rare earth-iron alloys},\ }\href
	{https://doi.org/10.1088/0022-3727/27/1/001} {\bibfield  {journal} {\bibinfo
			{journal} {J. Phys. D: Appl. Phys.}\ }\textbf {\bibinfo {volume} {27}},\
		\bibinfo {pages} {1} (\bibinfo {year} {1994})}\BibitemShut {NoStop}%
	\bibitem [{\citenamefont {Song}\ \emph {et~al.}(2021)\citenamefont {Song},
		\citenamefont {Shi}, \citenamefont {Deng}, \citenamefont {Xing},\ and\
		\citenamefont {Chen}}]{song;pms21}%
	\BibitemOpen
	\bibfield  {author} {\bibinfo {author} {\bibfnamefont {Y.}~\bibnamefont
			{Song}}, \bibinfo {author} {\bibfnamefont {N.}~\bibnamefont {Shi}}, \bibinfo
		{author} {\bibfnamefont {S.}~\bibnamefont {Deng}}, \bibinfo {author}
		{\bibfnamefont {X.}~\bibnamefont {Xing}},\ and\ \bibinfo {author}
		{\bibfnamefont {J.}~\bibnamefont {Chen}},\ }\bibfield  {title} {\bibinfo
		{title} {Negative thermal expansion in magnetic materials},\ }\href
	{https://doi.org/https://doi.org/10.1016/j.pmatsci.2021.100835} {\bibfield
		{journal} {\bibinfo  {journal} {Prog. Mater. Sci.}\ }\textbf {\bibinfo
			{volume} {121}},\ \bibinfo {pages} {100835} (\bibinfo {year}
		{2021})}\BibitemShut {NoStop}%
	\bibitem [{\citenamefont {Bandyopadhyay}\ \emph {et~al.}(2021)\citenamefont
		{Bandyopadhyay}, \citenamefont {Atulasimha},\ and\ \citenamefont
		{Barman}}]{bandy;aprev21}%
	\BibitemOpen
	\bibfield  {author} {\bibinfo {author} {\bibfnamefont {S.}~\bibnamefont
			{Bandyopadhyay}}, \bibinfo {author} {\bibfnamefont {J.}~\bibnamefont
			{Atulasimha}},\ and\ \bibinfo {author} {\bibfnamefont {A.}~\bibnamefont
			{Barman}},\ }\bibfield  {title} {\bibinfo {title} {{Magnetic straintronics:
				Manipulating the magnetization of magnetostrictive nanomagnets with strain
				for energy-efficient applications}},\ }\href
	{https://doi.org/10.1063/5.0062993} {\bibfield  {journal} {\bibinfo
			{journal} {Appl. Phys. Rev.}\ }\textbf {\bibinfo {volume} {8}},\ \bibinfo
		{pages} {041323} (\bibinfo {year} {2021})},\ \Eprint
	{https://arxiv.org/abs/https://doi.org/10.1063/5.0062993}
	{https://doi.org/10.1063/5.0062993} \BibitemShut {NoStop}%
	\bibitem [{\citenamefont {Sander}(2020)}]{sande;b;hmmm20}%
	\BibitemOpen
	\bibfield  {author} {\bibinfo {author} {\bibfnamefont {D.}~\bibnamefont
			{Sander}},\ }\bibinfo {title} {{Magnetostriction and Magnetoelasticity}},\
	in\ \href {https://doi.org/10.1007/978-3-030-63101-7_11-1} {\emph {\bibinfo
			{booktitle} {{Handbook of Magnetism and Magnetic Materials}}}},\ \bibinfo
	{editor} {edited by\ \bibinfo {editor} {\bibfnamefont {M.}~\bibnamefont
			{Coey}}\ and\ \bibinfo {editor} {\bibfnamefont {S.}~\bibnamefont {Parkin}}}\
	(\bibinfo  {publisher} {Springer International Publishing},\ \bibinfo
	{address} {Cham},\ \bibinfo {year} {2020})\ pp.\ \bibinfo {pages}
	{1--45}\BibitemShut {NoStop}%
	\bibitem [{\citenamefont {Doerr}\ \emph {et~al.}(2005)\citenamefont {Doerr},
		\citenamefont {Rotter},\ and\ \citenamefont {Lindbaum}}]{doerr;advp05}%
	\BibitemOpen
	\bibfield  {author} {\bibinfo {author} {\bibfnamefont {M.}~\bibnamefont
			{Doerr}}, \bibinfo {author} {\bibfnamefont {M.}~\bibnamefont {Rotter}},\ and\
		\bibinfo {author} {\bibfnamefont {A.}~\bibnamefont {Lindbaum}},\ }\bibfield
	{title} {\bibinfo {title} {{Magnetostriction in rare-earth based
				antiferromagnets}},\ }\href {https://doi.org/10.1080/00018730500037264}
	{\bibfield  {journal} {\bibinfo  {journal} {Adv. Phys.}\ }\textbf {\bibinfo
			{volume} {54}},\ \bibinfo {pages} {1} (\bibinfo {year} {2005})}\BibitemShut
	{NoStop}%
	\bibitem [{\citenamefont {Charilaou}\ \emph {et~al.}(2012)\citenamefont
		{Charilaou}, \citenamefont {Sheptyakov}, \citenamefont {L\"offler},\ and\
		\citenamefont {Gehring}}]{chari;prb12}%
	\BibitemOpen
	\bibfield  {author} {\bibinfo {author} {\bibfnamefont {M.}~\bibnamefont
			{Charilaou}}, \bibinfo {author} {\bibfnamefont {D.}~\bibnamefont
			{Sheptyakov}}, \bibinfo {author} {\bibfnamefont {J.~F.}\ \bibnamefont
			{L\"offler}},\ and\ \bibinfo {author} {\bibfnamefont {A.~U.}\ \bibnamefont
			{Gehring}},\ }\bibfield  {title} {\bibinfo {title} {{Large spontaneous
				magnetostriction in FeTiO$_3$ and adjustable magnetic configuration in
				Fe(III)-doped FeTiO$_3$}},\ }\href
	{https://doi.org/10.1103/PhysRevB.86.024439} {\bibfield  {journal} {\bibinfo
			{journal} {Phys. Rev. B}\ }\textbf {\bibinfo {volume} {86}},\ \bibinfo
		{pages} {024439} (\bibinfo {year} {2012})}\BibitemShut {NoStop}%
	\bibitem [{\citenamefont {Singh}\ \emph {et~al.}(2021)\citenamefont {Singh},
		\citenamefont {Samathrakis}, \citenamefont {Fortunato}, \citenamefont
		{Zemen}, \citenamefont {Shen}, \citenamefont {Gutfleisch},\ and\
		\citenamefont {Zhang}}]{singh;npjcm21}%
	\BibitemOpen
	\bibfield  {author} {\bibinfo {author} {\bibfnamefont {H.~K.}\ \bibnamefont
			{Singh}}, \bibinfo {author} {\bibfnamefont {I.}~\bibnamefont {Samathrakis}},
		\bibinfo {author} {\bibfnamefont {N.~M.}\ \bibnamefont {Fortunato}}, \bibinfo
		{author} {\bibfnamefont {J.}~\bibnamefont {Zemen}}, \bibinfo {author}
		{\bibfnamefont {C.}~\bibnamefont {Shen}}, \bibinfo {author} {\bibfnamefont
			{O.}~\bibnamefont {Gutfleisch}},\ and\ \bibinfo {author} {\bibfnamefont
			{H.}~\bibnamefont {Zhang}},\ }\bibfield  {title} {\bibinfo {title}
		{Multifunctional antiperovskites driven by strong magnetostructural
			coupling},\ }\href@noop {} {\bibfield  {journal} {\bibinfo  {journal} {npj
				Comput. Mater.}\ }\textbf {\bibinfo {volume} {7}},\ \bibinfo {pages} {1}
		(\bibinfo {year} {2021})}\BibitemShut {NoStop}%
	\bibitem [{\citenamefont {Miao}\ \emph {et~al.}(2021)\citenamefont {Miao},
		\citenamefont {Tan}, \citenamefont {Lee}, \citenamefont {Ishikawa},
		\citenamefont {Torii}, \citenamefont {Yonemura}, \citenamefont {Koda},
		\citenamefont {Komatsu}, \citenamefont {Machida}, \citenamefont
		{Sano-Furukawa}, \citenamefont {Hattori}, \citenamefont {Lin}, \citenamefont
		{Li}, \citenamefont {Mochiku}, \citenamefont {Kikuchi}, \citenamefont
		{Kawashima}, \citenamefont {Takahashi}, \citenamefont {Huang}, \citenamefont
		{Itoh}, \citenamefont {Kadono}, \citenamefont {Wang}, \citenamefont {Pan},
		\citenamefont {Yamauchi},\ and\ \citenamefont {Kamiyama}}]{miao;prb21}%
	\BibitemOpen
	\bibfield  {author} {\bibinfo {author} {\bibfnamefont {P.}~\bibnamefont
			{Miao}}, \bibinfo {author} {\bibfnamefont {Z.}~\bibnamefont {Tan}}, \bibinfo
		{author} {\bibfnamefont {S.}~\bibnamefont {Lee}}, \bibinfo {author}
		{\bibfnamefont {Y.}~\bibnamefont {Ishikawa}}, \bibinfo {author}
		{\bibfnamefont {S.}~\bibnamefont {Torii}}, \bibinfo {author} {\bibfnamefont
			{M.}~\bibnamefont {Yonemura}}, \bibinfo {author} {\bibfnamefont
			{A.}~\bibnamefont {Koda}}, \bibinfo {author} {\bibfnamefont {K.}~\bibnamefont
			{Komatsu}}, \bibinfo {author} {\bibfnamefont {S.}~\bibnamefont {Machida}},
		\bibinfo {author} {\bibfnamefont {A.}~\bibnamefont {Sano-Furukawa}}, \bibinfo
		{author} {\bibfnamefont {T.}~\bibnamefont {Hattori}}, \bibinfo {author}
		{\bibfnamefont {X.}~\bibnamefont {Lin}}, \bibinfo {author} {\bibfnamefont
			{K.}~\bibnamefont {Li}}, \bibinfo {author} {\bibfnamefont {T.}~\bibnamefont
			{Mochiku}}, \bibinfo {author} {\bibfnamefont {R.}~\bibnamefont {Kikuchi}},
		\bibinfo {author} {\bibfnamefont {C.}~\bibnamefont {Kawashima}}, \bibinfo
		{author} {\bibfnamefont {H.}~\bibnamefont {Takahashi}}, \bibinfo {author}
		{\bibfnamefont {Q.}~\bibnamefont {Huang}}, \bibinfo {author} {\bibfnamefont
			{S.}~\bibnamefont {Itoh}}, \bibinfo {author} {\bibfnamefont {R.}~\bibnamefont
			{Kadono}}, \bibinfo {author} {\bibfnamefont {Y.}~\bibnamefont {Wang}},
		\bibinfo {author} {\bibfnamefont {F.}~\bibnamefont {Pan}}, \bibinfo {author}
		{\bibfnamefont {K.}~\bibnamefont {Yamauchi}},\ and\ \bibinfo {author}
		{\bibfnamefont {T.}~\bibnamefont {Kamiyama}},\ }\bibfield  {title} {\bibinfo
		{title} {Origin of magnetovolume effect in a cobaltite},\ }\href
	{https://doi.org/10.1103/PhysRevB.103.094302} {\bibfield  {journal} {\bibinfo
			{journal} {Phys. Rev. B}\ }\textbf {\bibinfo {volume} {103}},\ \bibinfo
		{pages} {094302} (\bibinfo {year} {2021})}\BibitemShut {NoStop}%
	\bibitem [{\citenamefont {Casillas-Trujillo}\ \emph {et~al.}(2021)\citenamefont
		{Casillas-Trujillo}, \citenamefont {Armiento},\ and\ \citenamefont
		{Alling}}]{casil;prm21}%
	\BibitemOpen
	\bibfield  {author} {\bibinfo {author} {\bibfnamefont {L.}~\bibnamefont
			{Casillas-Trujillo}}, \bibinfo {author} {\bibfnamefont {R.}~\bibnamefont
			{Armiento}},\ and\ \bibinfo {author} {\bibfnamefont {B.}~\bibnamefont
			{Alling}},\ }\bibfield  {title} {\bibinfo {title} {Identification of
			materials with strong magnetostructural coupling using computational
			high-throughput screening},\ }\href
	{https://doi.org/10.1103/PhysRevMaterials.5.034417} {\bibfield  {journal}
		{\bibinfo  {journal} {Phys. Rev. Materials}\ }\textbf {\bibinfo {volume}
			{5}},\ \bibinfo {pages} {034417} (\bibinfo {year} {2021})}\BibitemShut
	{NoStop}%
	\bibitem [{\citenamefont {Dey}\ \emph {et~al.}(2021)\citenamefont {Dey},
		\citenamefont {Sauerland}, \citenamefont {Ouladdiaf}, \citenamefont
		{Beauvois}, \citenamefont {Wadepohl},\ and\ \citenamefont
		{Klingeler}}]{dey;prb21}%
	\BibitemOpen
	\bibfield  {author} {\bibinfo {author} {\bibfnamefont {K.}~\bibnamefont
			{Dey}}, \bibinfo {author} {\bibfnamefont {S.}~\bibnamefont {Sauerland}},
		\bibinfo {author} {\bibfnamefont {B.}~\bibnamefont {Ouladdiaf}}, \bibinfo
		{author} {\bibfnamefont {K.}~\bibnamefont {Beauvois}}, \bibinfo {author}
		{\bibfnamefont {H.}~\bibnamefont {Wadepohl}},\ and\ \bibinfo {author}
		{\bibfnamefont {R.}~\bibnamefont {Klingeler}},\ }\bibfield  {title} {\bibinfo
		{title} {{Magnetostructural coupling in ilmenite-type NiTiO$_3$}},\ }\href
	{https://doi.org/10.1103/PhysRevB.103.134438} {\bibfield  {journal} {\bibinfo
			{journal} {Phys. Rev. B}\ }\textbf {\bibinfo {volume} {103}},\ \bibinfo
		{pages} {134438} (\bibinfo {year} {2021})}\BibitemShut {NoStop}%
	\bibitem [{\citenamefont {Prescher}\ and\ \citenamefont
		{Prakapenka}(2015)}]{presc;hpr15}%
	\BibitemOpen
	\bibfield  {author} {\bibinfo {author} {\bibfnamefont {C.}~\bibnamefont
			{Prescher}}\ and\ \bibinfo {author} {\bibfnamefont {V.~B.}\ \bibnamefont
			{Prakapenka}},\ }\bibfield  {title} {\bibinfo {title} {{DIOPTAS: a program
				for reduction of two-dimensional X-ray diffraction data and data
				exploration}},\ }\href {https://doi.org/10.1080/08957959.2015.1059835}
	{\bibfield  {journal} {\bibinfo  {journal} {High Pressure Res.}\ }\textbf
		{\bibinfo {volume} {35}},\ \bibinfo {pages} {223} (\bibinfo {year}
		{2015})}\BibitemShut {NoStop}%
	\bibitem [{\citenamefont {Yang}\ \emph {et~al.}(2015)\citenamefont {Yang},
		\citenamefont {Juh\'{a}s}, \citenamefont {Farrow},\ and\ \citenamefont
		{Billinge}}]{yang;arxiv15}%
	\BibitemOpen
	\bibfield  {author} {\bibinfo {author} {\bibfnamefont {X.}~\bibnamefont
			{Yang}}, \bibinfo {author} {\bibfnamefont {P.}~\bibnamefont {Juh\'{a}s}},
		\bibinfo {author} {\bibfnamefont {C.}~\bibnamefont {Farrow}},\ and\ \bibinfo
		{author} {\bibfnamefont {S.~J.~L.}\ \bibnamefont {Billinge}},\ }\bibfield
	{title} {\bibinfo {title} {x{PDF}suite: an end-to-end software solution for
			high throughput pair distribution function transformation, visualization and
			analysis},\ }\href {http://arxiv.org/abs/1402.3163} {\bibfield  {journal}
		{\bibinfo  {journal} {arXiv}\ } (\bibinfo {year} {2015})},\ \bibinfo {note}
	{1402.3163}\BibitemShut {NoStop}%
	\bibitem [{\citenamefont {Egami}\ and\ \citenamefont
		{Billinge}(2012)}]{egami;b;utbp12}%
	\BibitemOpen
	\bibfield  {author} {\bibinfo {author} {\bibfnamefont {T.}~\bibnamefont
			{Egami}}\ and\ \bibinfo {author} {\bibfnamefont {S.~J.~L.}\ \bibnamefont
			{Billinge}},\ }\href
	{http://store.elsevier.com/product.jsp?lid=0\&iid=73\&sid=0\&isbn=9780080971414}
	{\emph {\bibinfo {title} {Underneath the Bragg peaks: structural analysis of
				complex materials}}},\ \bibinfo {edition} {2nd}\ ed.\ (\bibinfo  {publisher}
	{Elsevier},\ \bibinfo {address} {Amsterdam},\ \bibinfo {year}
	{2012})\BibitemShut {NoStop}%
	\bibitem [{\citenamefont {Farrow}\ \emph {et~al.}(2007)\citenamefont {Farrow},
		\citenamefont {Juh\'as}, \citenamefont {Liu}, \citenamefont {Bryndin},
		\citenamefont {{Bo\v zin}}, \citenamefont {Bloch}, \citenamefont {Proffen},\
		and\ \citenamefont {Billinge}}]{farro;jpcm07}%
	\BibitemOpen
	\bibfield  {author} {\bibinfo {author} {\bibfnamefont {C.~L.}\ \bibnamefont
			{Farrow}}, \bibinfo {author} {\bibfnamefont {P.}~\bibnamefont {Juh\'as}},
		\bibinfo {author} {\bibfnamefont {J.}~\bibnamefont {Liu}}, \bibinfo {author}
		{\bibfnamefont {D.}~\bibnamefont {Bryndin}}, \bibinfo {author} {\bibfnamefont
			{E.~S.}\ \bibnamefont {{Bo\v zin}}}, \bibinfo {author} {\bibfnamefont
			{J.}~\bibnamefont {Bloch}}, \bibinfo {author} {\bibfnamefont
			{T.}~\bibnamefont {Proffen}},\ and\ \bibinfo {author} {\bibfnamefont
			{S.~J.~L.}\ \bibnamefont {Billinge}},\ }\bibfield  {title} {\bibinfo {title}
		{{PDFfit2} and {PDFgui}: Computer programs for studying nanostructure in
			crystals},\ }\href {https://doi.org/10.1088/0953-8984/19/33/335219}
	{\bibfield  {journal} {\bibinfo  {journal} {J. Phys.: Condens. Mat.}\
		}\textbf {\bibinfo {volume} {19}},\ \bibinfo {pages} {335219} (\bibinfo
		{year} {2007})}\BibitemShut {NoStop}%
	\bibitem [{\citenamefont {Campbell}\ \emph {et~al.}(2006)\citenamefont
		{Campbell}, \citenamefont {Stokes}, \citenamefont {Tanner},\ and\
		\citenamefont {Hatch}}]{campb;jac06}%
	\BibitemOpen
	\bibfield  {author} {\bibinfo {author} {\bibfnamefont {B.}~\bibnamefont
			{Campbell}}, \bibinfo {author} {\bibfnamefont {H.}~\bibnamefont {Stokes}},
		\bibinfo {author} {\bibfnamefont {D.}~\bibnamefont {Tanner}},\ and\ \bibinfo
		{author} {\bibfnamefont {D.}~\bibnamefont {Hatch}},\ }\bibfield  {title}
	{\bibinfo {title} {{ISODISPLACE}: {A}n {I}nternet {T}ool for {E}xploring
			{S}tructural {D}istortions},\ }\href@noop {} {\bibfield  {journal} {\bibinfo
			{journal} {J. Appl. Crystallogr.}\ }\textbf {\bibinfo {volume} {39}},\
		\bibinfo {pages} {607} (\bibinfo {year} {2006})}\BibitemShut {NoStop}%
	\bibitem [{\citenamefont {Stokes}\ \emph
		{et~al.}(2023{\natexlab{a}})\citenamefont {Stokes}, \citenamefont {Hatch},\
		and\ \citenamefont {Campbell}}]{stoke;webIso}%
	\BibitemOpen
	\bibfield  {author} {\bibinfo {author} {\bibfnamefont {H.~T.}\ \bibnamefont
			{Stokes}}, \bibinfo {author} {\bibfnamefont {D.~M.}\ \bibnamefont {Hatch}},\
		and\ \bibinfo {author} {\bibfnamefont {B.~J.}\ \bibnamefont {Campbell}},\
	}\href {https://iso.byu.edu} {\bibinfo {title} {{ISODISTORT, ISOTROPY
				Software Suite}}} (\bibinfo {year} {2023}{\natexlab{a}})\BibitemShut
	{NoStop}%
	\bibitem [{\citenamefont {Hatch}\ and\ \citenamefont
		{Stokes}(2003)}]{hatch;jac03}%
	\BibitemOpen
	\bibfield  {author} {\bibinfo {author} {\bibfnamefont {D.~M.}\ \bibnamefont
			{Hatch}}\ and\ \bibinfo {author} {\bibfnamefont {H.~T.}\ \bibnamefont
			{Stokes}},\ }\bibfield  {title} {\bibinfo {title} {{{\it INVARIANTS}: program
				for obtaining a list of invariant polynomials of the order-parameter
				components associated with irreducible representations of a space group}},\
	}\href {https://doi.org/10.1107/S0021889803005946} {\bibfield  {journal}
		{\bibinfo  {journal} {J. Appl. Crystallogr.}\ }\textbf {\bibinfo {volume}
			{36}},\ \bibinfo {pages} {951} (\bibinfo {year} {2003})}\BibitemShut
	{NoStop}%
	\bibitem [{\citenamefont {Stokes}\ \emph
		{et~al.}(2023{\natexlab{b}})\citenamefont {Stokes}, \citenamefont {Hatch},\
		and\ \citenamefont {Campbell}}]{stoke;webInv}%
	\BibitemOpen
	\bibfield  {author} {\bibinfo {author} {\bibfnamefont {H.~T.}\ \bibnamefont
			{Stokes}}, \bibinfo {author} {\bibfnamefont {D.~M.}\ \bibnamefont {Hatch}},\
		and\ \bibinfo {author} {\bibfnamefont {B.~J.}\ \bibnamefont {Campbell}},\
	}\href {https://iso.byu.edu/} {\bibinfo {title} {{ISODISTORT, ISOTROPY
				Software Suite}}} (\bibinfo {year} {2023}{\natexlab{b}})\BibitemShut
	{NoStop}%
	\bibitem [{\citenamefont {Chatterji}\ \emph {et~al.}(2010)\citenamefont
		{Chatterji}, \citenamefont {Iles}, \citenamefont {Ouladdiaf},\ and\
		\citenamefont {Hansen}}]{chatt;jpcm10}%
	\BibitemOpen
	\bibfield  {author} {\bibinfo {author} {\bibfnamefont {T.}~\bibnamefont
			{Chatterji}}, \bibinfo {author} {\bibfnamefont {G.~N.}\ \bibnamefont {Iles}},
		\bibinfo {author} {\bibfnamefont {B.}~\bibnamefont {Ouladdiaf}},\ and\
		\bibinfo {author} {\bibfnamefont {T.~C.}\ \bibnamefont {Hansen}},\ }\bibfield
	{title} {\bibinfo {title} {{Magnetoelastic effect in MF$_2$ (M = Mn, Fe, Ni)
				investigated by neutron powder diffraction}},\ }\href
	{https://doi.org/10.1088/0953-8984/22/31/316001} {\bibfield  {journal}
		{\bibinfo  {journal} {J. Phys.: Condens. Mat.}\ }\textbf {\bibinfo {volume}
			{22}},\ \bibinfo {pages} {316001} (\bibinfo {year} {2010})}\BibitemShut
	{NoStop}%
	\bibitem [{\citenamefont {Chatterji}\ and\ \citenamefont
		{Hansen}(2011)}]{chatt;jpcm11}%
	\BibitemOpen
	\bibfield  {author} {\bibinfo {author} {\bibfnamefont {T.}~\bibnamefont
			{Chatterji}}\ and\ \bibinfo {author} {\bibfnamefont {T.~C.}\ \bibnamefont
			{Hansen}},\ }\bibfield  {title} {\bibinfo {title} {{Magnetoelastic effects in
				Jahn--Teller distorted CrF$_2$ and CuF$_2$ studied by neutron powder
				diffraction}},\ }\href@noop {} {\bibfield  {journal} {\bibinfo  {journal} {J.
				Phys.: Condens. Mat.}\ }\textbf {\bibinfo {volume} {23}},\ \bibinfo {pages}
		{276007} (\bibinfo {year} {2011})}\BibitemShut {NoStop}%
	\bibitem [{\citenamefont {Paszkowicz}\ and\ \citenamefont
		{Dynowska}(1997)}]{paszk;appa97}%
	\BibitemOpen
	\bibfield  {author} {\bibinfo {author} {\bibfnamefont {W.}~\bibnamefont
			{Paszkowicz}}\ and\ \bibinfo {author} {\bibfnamefont {E.}~\bibnamefont
			{Dynowska}},\ }\bibfield  {title} {\bibinfo {title} {{HIGH PRESSURE - HIGH
				TEMPERATURE DIFFRACTION STUDY OF MnTe USING SYNCHROTRON RADIATION}},\
	}\href@noop {} {\bibfield  {journal} {\bibinfo  {journal} {Acta Phys. Pol.
				A}\ }\textbf {\bibinfo {volume} {91}},\ \bibinfo {pages} {939} (\bibinfo
		{year} {1997})}\BibitemShut {NoStop}%
	\bibitem [{\citenamefont {Frandsen}\ \emph {et~al.}(2014)\citenamefont
		{Frandsen}, \citenamefont {Yang},\ and\ \citenamefont
		{Billinge}}]{frand;aca14}%
	\BibitemOpen
	\bibfield  {author} {\bibinfo {author} {\bibfnamefont {B.~A.}\ \bibnamefont
			{Frandsen}}, \bibinfo {author} {\bibfnamefont {X.}~\bibnamefont {Yang}},\
		and\ \bibinfo {author} {\bibfnamefont {S.~J.~L.}\ \bibnamefont {Billinge}},\
	}\bibfield  {title} {\bibinfo {title} {Magnetic pair distribution function
			analysis of local magnetic correlations},\ }\href
	{https://doi.org/10.1107/S2053273313033081} {\bibfield  {journal} {\bibinfo
			{journal} {Acta Crystallogr. A}\ }\textbf {\bibinfo {volume} {70}},\ \bibinfo
		{pages} {3} (\bibinfo {year} {2014})}\BibitemShut {NoStop}%
	\bibitem [{\citenamefont {Frandsen}\ and\ \citenamefont
		{Billinge}(2015)}]{frand;aca15}%
	\BibitemOpen
	\bibfield  {author} {\bibinfo {author} {\bibfnamefont {B.~A.}\ \bibnamefont
			{Frandsen}}\ and\ \bibinfo {author} {\bibfnamefont {S.~J.~L.}\ \bibnamefont
			{Billinge}},\ }\bibfield  {title} {\bibinfo {title} {Magnetic structure
			determination from the magnetic pair distribution function {(mPDF)}: ground
			state of {MnO}},\ }\href {https://doi.org/10.1107/S205327331500306X}
	{\bibfield  {journal} {\bibinfo  {journal} {Acta Crystallogr. A}\ }\textbf
		{\bibinfo {volume} {71}},\ \bibinfo {pages} {325} (\bibinfo {year}
		{2015})}\BibitemShut {NoStop}%
	\bibitem [{\citenamefont {Mu}\ \emph {et~al.}(2019)\citenamefont {Mu},
		\citenamefont {Hermann}, \citenamefont {Gorsse}, \citenamefont {Zhao},
		\citenamefont {Manley}, \citenamefont {Fishman},\ and\ \citenamefont
		{Lindsay}}]{mu;prm19}%
	\BibitemOpen
	\bibfield  {author} {\bibinfo {author} {\bibfnamefont {S.}~\bibnamefont
			{Mu}}, \bibinfo {author} {\bibfnamefont {R.~P.}\ \bibnamefont {Hermann}},
		\bibinfo {author} {\bibfnamefont {S.}~\bibnamefont {Gorsse}}, \bibinfo
		{author} {\bibfnamefont {H.}~\bibnamefont {Zhao}}, \bibinfo {author}
		{\bibfnamefont {M.~E.}\ \bibnamefont {Manley}}, \bibinfo {author}
		{\bibfnamefont {R.~S.}\ \bibnamefont {Fishman}},\ and\ \bibinfo {author}
		{\bibfnamefont {L.}~\bibnamefont {Lindsay}},\ }\bibfield  {title} {\bibinfo
		{title} {{Phonons, magnons, and lattice thermal transport in
				antiferromagnetic semiconductor MnTe}},\ }\href
	{https://doi.org/10.1103/PhysRevMaterials.3.025403} {\bibfield  {journal}
		{\bibinfo  {journal} {Phys. Rev. Materials}\ }\textbf {\bibinfo {volume}
			{3}},\ \bibinfo {pages} {025403} (\bibinfo {year} {2019})}\BibitemShut
	{NoStop}%
	\bibitem [{\citenamefont {Szuszkiewicz}\ \emph {et~al.}(2006)\citenamefont
		{Szuszkiewicz}, \citenamefont {Dynowska}, \citenamefont {Witkowska},\ and\
		\citenamefont {Hennion}}]{szusz;prb06}%
	\BibitemOpen
	\bibfield  {author} {\bibinfo {author} {\bibfnamefont {W.}~\bibnamefont
			{Szuszkiewicz}}, \bibinfo {author} {\bibfnamefont {E.}~\bibnamefont
			{Dynowska}}, \bibinfo {author} {\bibfnamefont {B.}~\bibnamefont
			{Witkowska}},\ and\ \bibinfo {author} {\bibfnamefont {B.}~\bibnamefont
			{Hennion}},\ }\bibfield  {title} {\bibinfo {title} {{Spin-wave measurements
				on hexagonal MnTe NiAs-type structure by inelastic neutron scattering}},\
	}\href {https://doi.org/10.1103/PhysRevB.73.104403} {\bibfield  {journal}
		{\bibinfo  {journal} {Phys. Rev. B}\ }\textbf {\bibinfo {volume} {73}},\
		\bibinfo {pages} {104403} (\bibinfo {year} {2006})}\BibitemShut {NoStop}%
	\bibitem [{\citenamefont {Zapf}\ \emph {et~al.}(2008)\citenamefont {Zapf},
		\citenamefont {Correa}, \citenamefont {Sengupta}, \citenamefont {Batista},
		\citenamefont {Tsukamoto}, \citenamefont {Kawashima}, \citenamefont {Egan},
		\citenamefont {Pantea}, \citenamefont {Migliori}, \citenamefont {Betts},
		\citenamefont {Jaime},\ and\ \citenamefont {Paduan-Filho}}]{zapf;prb08}%
	\BibitemOpen
	\bibfield  {author} {\bibinfo {author} {\bibfnamefont {V.~S.}\ \bibnamefont
			{Zapf}}, \bibinfo {author} {\bibfnamefont {V.~F.}\ \bibnamefont {Correa}},
		\bibinfo {author} {\bibfnamefont {P.}~\bibnamefont {Sengupta}}, \bibinfo
		{author} {\bibfnamefont {C.~D.}\ \bibnamefont {Batista}}, \bibinfo {author}
		{\bibfnamefont {M.}~\bibnamefont {Tsukamoto}}, \bibinfo {author}
		{\bibfnamefont {N.}~\bibnamefont {Kawashima}}, \bibinfo {author}
		{\bibfnamefont {P.}~\bibnamefont {Egan}}, \bibinfo {author} {\bibfnamefont
			{C.}~\bibnamefont {Pantea}}, \bibinfo {author} {\bibfnamefont
			{A.}~\bibnamefont {Migliori}}, \bibinfo {author} {\bibfnamefont {J.~B.}\
			\bibnamefont {Betts}}, \bibinfo {author} {\bibfnamefont {M.}~\bibnamefont
			{Jaime}},\ and\ \bibinfo {author} {\bibfnamefont {A.}~\bibnamefont
			{Paduan-Filho}},\ }\bibfield  {title} {\bibinfo {title} {Direct measurement
			of spin correlations using magnetostriction},\ }\href
	{https://doi.org/10.1103/PhysRevB.77.020404} {\bibfield  {journal} {\bibinfo
			{journal} {Phys. Rev. B}\ }\textbf {\bibinfo {volume} {77}},\ \bibinfo
		{pages} {020404} (\bibinfo {year} {2008})}\BibitemShut {NoStop}%
	\bibitem [{\citenamefont {Bloch}\ and\ \citenamefont
		{Maury}(1973)}]{bloch;prb73}%
	\BibitemOpen
	\bibfield  {author} {\bibinfo {author} {\bibfnamefont {D.}~\bibnamefont
			{Bloch}}\ and\ \bibinfo {author} {\bibfnamefont {R.}~\bibnamefont {Maury}},\
	}\bibfield  {title} {\bibinfo {title} {{Uniaxial Stress Experiments and
				Magnetoelastic Interactions in Manganese Oxide}},\ }\href
	{https://doi.org/10.1103/PhysRevB.7.4883} {\bibfield  {journal} {\bibinfo
			{journal} {Phys. Rev. B}\ }\textbf {\bibinfo {volume} {7}},\ \bibinfo {pages}
		{4883} (\bibinfo {year} {1973})}\BibitemShut {NoStop}%
	\bibitem [{\citenamefont {Oey}\ \emph {et~al.}(2020)\citenamefont {Oey},
		\citenamefont {Bocarsly}, \citenamefont {Mann}, \citenamefont {Levin},
		\citenamefont {Shatruk},\ and\ \citenamefont {Seshadri}}]{oey;apl20}%
	\BibitemOpen
	\bibfield  {author} {\bibinfo {author} {\bibfnamefont {Y.~M.}\ \bibnamefont
			{Oey}}, \bibinfo {author} {\bibfnamefont {J.~D.}\ \bibnamefont {Bocarsly}},
		\bibinfo {author} {\bibfnamefont {D.}~\bibnamefont {Mann}}, \bibinfo {author}
		{\bibfnamefont {E.~E.}\ \bibnamefont {Levin}}, \bibinfo {author}
		{\bibfnamefont {M.}~\bibnamefont {Shatruk}},\ and\ \bibinfo {author}
		{\bibfnamefont {R.}~\bibnamefont {Seshadri}},\ }\bibfield  {title} {\bibinfo
		{title} {{Structural changes upon magnetic ordering in magnetocaloric
				AlFe$_2$B$_2$}},\ }\href {https://doi.org/10.1063/5.0007266} {\bibfield
		{journal} {\bibinfo  {journal} {Appl. Phys. Lett.}\ }\textbf {\bibinfo
			{volume} {116}},\ \bibinfo {pages} {212403} (\bibinfo {year} {2020})},\
	\Eprint {https://arxiv.org/abs/https://doi.org/10.1063/5.0007266}
	{https://doi.org/10.1063/5.0007266} \BibitemShut {NoStop}%
	\bibitem [{\citenamefont {Pascard}\ and\ \citenamefont
		{Globus}(1982)}]{pasca;jap82}%
	\BibitemOpen
	\bibfield  {author} {\bibinfo {author} {\bibfnamefont {H.}~\bibnamefont
			{Pascard}}\ and\ \bibinfo {author} {\bibfnamefont {A.}~\bibnamefont
			{Globus}},\ }\bibfield  {title} {\bibinfo {title} {Exchange striction and
			crystal lattice in domains and domain walls},\ }\href
	{https://doi.org/10.1063/1.330833} {\bibfield  {journal} {\bibinfo  {journal}
			{J. Appl. Phys.}\ }\textbf {\bibinfo {volume} {53}},\ \bibinfo {pages} {2425}
		(\bibinfo {year} {1982})},\ \Eprint
	{https://arxiv.org/abs/https://doi.org/10.1063/1.330833}
	{https://doi.org/10.1063/1.330833} \BibitemShut {NoStop}%
	\bibitem [{\citenamefont {Carpenter}\ \emph {et~al.}(2012)\citenamefont
		{Carpenter}, \citenamefont {Zhang},\ and\ \citenamefont
		{Howard}}]{carpe;jpcm12}%
	\BibitemOpen
	\bibfield  {author} {\bibinfo {author} {\bibfnamefont {M.~A.}\ \bibnamefont
			{Carpenter}}, \bibinfo {author} {\bibfnamefont {Z.}~\bibnamefont {Zhang}},\
		and\ \bibinfo {author} {\bibfnamefont {C.~J.}\ \bibnamefont {Howard}},\
	}\bibfield  {title} {\bibinfo {title} {{A linear-quadratic order parameter
				coupling model for magnetoelastic phase transitions in Fe$_{1-x}$O and
				MnO}},\ }\href {https://doi.org/10.1088/0953-8984/24/15/156002} {\bibfield
		{journal} {\bibinfo  {journal} {J. Phys.: Condens. Mat.}\ }\textbf {\bibinfo
			{volume} {24}},\ \bibinfo {pages} {156002} (\bibinfo {year}
		{2012})}\BibitemShut {NoStop}%
	\bibitem [{\citenamefont {Chatterji}\ \emph {et~al.}(2012)\citenamefont
		{Chatterji}, \citenamefont {Ouladdiaf}, \citenamefont {Henry},\ and\
		\citenamefont {Bhattacharya}}]{chatt;jpcm12}%
	\BibitemOpen
	\bibfield  {author} {\bibinfo {author} {\bibfnamefont {T.}~\bibnamefont
			{Chatterji}}, \bibinfo {author} {\bibfnamefont {B.}~\bibnamefont
			{Ouladdiaf}}, \bibinfo {author} {\bibfnamefont {P.~F.}\ \bibnamefont
			{Henry}},\ and\ \bibinfo {author} {\bibfnamefont {D.}~\bibnamefont
			{Bhattacharya}},\ }\bibfield  {title} {\bibinfo {title} {{Magnetoelastic
				effects in multiferroic YMnO$_3$}},\ }\href
	{https://doi.org/10.1088/0953-8984/24/33/336003} {\bibfield  {journal}
		{\bibinfo  {journal} {J. Phys.: Condens. Mat.}\ }\textbf {\bibinfo {volume}
			{24}},\ \bibinfo {pages} {336003} (\bibinfo {year} {2012})}\BibitemShut
	{NoStop}%
	\bibitem [{\citenamefont {Oravova}\ \emph {et~al.}(2013)\citenamefont
		{Oravova}, \citenamefont {Zhang}, \citenamefont {Church}, \citenamefont
		{Harrison}, \citenamefont {Howard},\ and\ \citenamefont
		{Carpenter}}]{oravo;jpcm13}%
	\BibitemOpen
	\bibfield  {author} {\bibinfo {author} {\bibfnamefont {L.}~\bibnamefont
			{Oravova}}, \bibinfo {author} {\bibfnamefont {Z.}~\bibnamefont {Zhang}},
		\bibinfo {author} {\bibfnamefont {N.}~\bibnamefont {Church}}, \bibinfo
		{author} {\bibfnamefont {R.~J.}\ \bibnamefont {Harrison}}, \bibinfo {author}
		{\bibfnamefont {C.~J.}\ \bibnamefont {Howard}},\ and\ \bibinfo {author}
		{\bibfnamefont {M.~A.}\ \bibnamefont {Carpenter}},\ }\bibfield  {title}
	{\bibinfo {title} {{Elastic and anelastic relaxations accompanying magnetic
				ordering and spin-flop transitions in hematite, Fe$_2$O$_3$}},\ }\href
	{https://doi.org/10.1088/0953-8984/25/11/116006} {\bibfield  {journal}
		{\bibinfo  {journal} {J. Phys.: Condens. Mat.}\ }\textbf {\bibinfo {volume}
			{25}},\ \bibinfo {pages} {116006} (\bibinfo {year} {2013})}\BibitemShut
	{NoStop}%
	\bibitem [{\citenamefont {Moseley}\ \emph {et~al.}(2022)\citenamefont
		{Moseley}, \citenamefont {Taddei}, \citenamefont {Yan}, \citenamefont
		{McGuire}, \citenamefont {Calder}, \citenamefont {Polash}, \citenamefont
		{Vashaee}, \citenamefont {Zhang}, \citenamefont {Zhao}, \citenamefont
		{Parker}, \citenamefont {Fishman},\ and\ \citenamefont
		{Hermann}}]{mosel;prm22}%
	\BibitemOpen
	\bibfield  {author} {\bibinfo {author} {\bibfnamefont {D.~H.}\ \bibnamefont
			{Moseley}}, \bibinfo {author} {\bibfnamefont {K.~M.}\ \bibnamefont {Taddei}},
		\bibinfo {author} {\bibfnamefont {J.}~\bibnamefont {Yan}}, \bibinfo {author}
		{\bibfnamefont {M.~A.}\ \bibnamefont {McGuire}}, \bibinfo {author}
		{\bibfnamefont {S.}~\bibnamefont {Calder}}, \bibinfo {author} {\bibfnamefont
			{M.~M.~H.}\ \bibnamefont {Polash}}, \bibinfo {author} {\bibfnamefont
			{D.}~\bibnamefont {Vashaee}}, \bibinfo {author} {\bibfnamefont
			{X.}~\bibnamefont {Zhang}}, \bibinfo {author} {\bibfnamefont
			{H.}~\bibnamefont {Zhao}}, \bibinfo {author} {\bibfnamefont {D.~S.}\
			\bibnamefont {Parker}}, \bibinfo {author} {\bibfnamefont {R.~S.}\
			\bibnamefont {Fishman}},\ and\ \bibinfo {author} {\bibfnamefont {R.~P.}\
			\bibnamefont {Hermann}},\ }\bibfield  {title} {\bibinfo {title} {{Giant
				doping response of magnetic anisotropy in MnTe}},\ }\href
	{https://doi.org/10.1103/PhysRevMaterials.6.014404} {\bibfield  {journal}
		{\bibinfo  {journal} {Phys. Rev. Materials}\ }\textbf {\bibinfo {volume}
			{6}},\ \bibinfo {pages} {014404} (\bibinfo {year} {2022})}\BibitemShut
	{NoStop}%
\end{thebibliography}

%

\end{document}


\title{
	Supplemental Information: Giant spontaneous magnetostriction in MnTe driven by a novel magnetostructural coupling mechanism
}
	
	\author{Raju Baral}
	\affiliation{ %
		Department of Physics and Astronomy, Brigham Young University, Provo, Utah 84602, USA.
	} %

	\author{Milinda Abeykoon}
	\affiliation{ %
		Photon Sciences Division, Brookhaven National Laboratory, Upton, NY, 11973 USA.
	} %
 
	\author{Branton J. Campbell}
	\affiliation{ %
		Department of Physics and Astronomy, Brigham Young University, Provo, Utah 84602, USA.
	} %

	\author{Benjamin A. Frandsen}
	\affiliation{ %
		Department of Physics and Astronomy, Brigham Young University, Provo, Utah 84602, USA.
	} %

\maketitle

The Supplemental Information includes additional details and data regarding the PDF fits, the Debye-Gr\"uneisen modeling of the lattice parameters and unit cell volume, the linear scaling between the lattice response and the local magnetic order parameter, the PDF measurements with an \textit{in situ} magnetic field, and the magnetic domain structure and associated free-energy arguments for Na-doped MnTe.

\begin{figure*}
	
 \centerline{\includegraphics[ width=150mm]{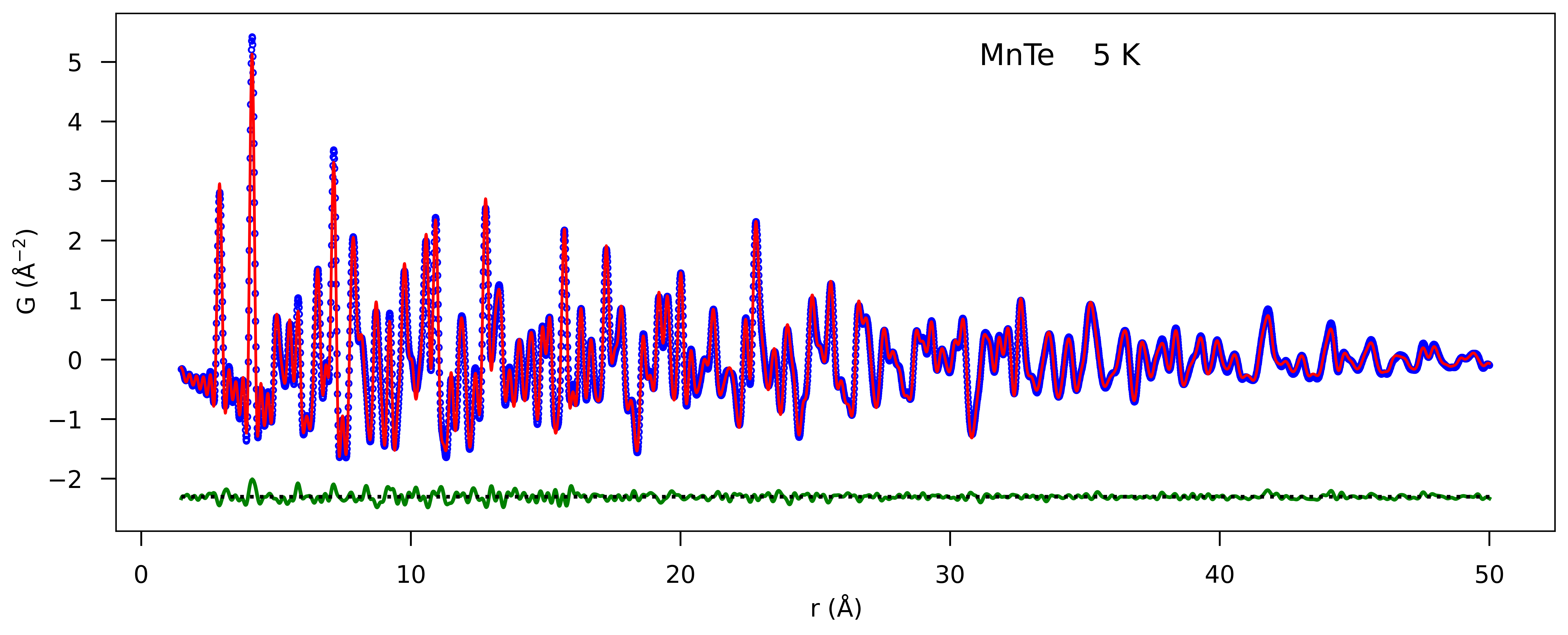}}
	\caption{\label{fig:MnTe-fit} X-ray PDF fit for pure MnTe at 5~K. The blue symbols represent the experimental data, the red curve the best-fit calculated PDF, and the green curve the fit residual, offset vertically for clarity.}
		
\end{figure*}

\begin{figure*}
	
 \centerline{\includegraphics[ width=150mm]{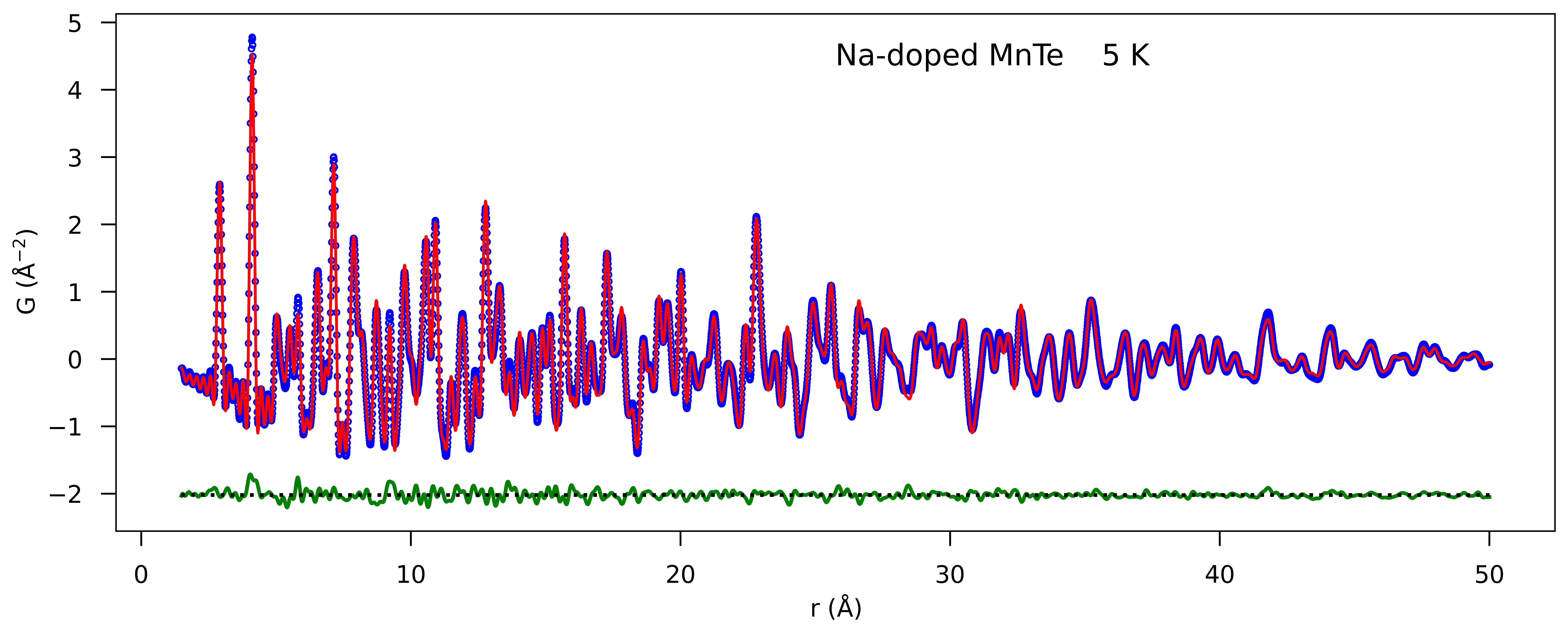}}
	\caption{\label{fig:Na-MnTe-fit} X-ray PDF fit for MnTe doped with 2\% Na at 5~K. The blue symbols represent the experimental data, the red curve the best-fit calculated PDF, and the green curve the fit residual, offset vertically for clarity.}

\end{figure*}

\textbf{Representative PDF fits.}
Representative fits to the x-ray PDF data collected at 5~K using the published hexagonal structure for MnTe are shown for pure MnTe in Fig.~\ref{fig:MnTe-fit}
and for Na-doped MnTe in Fig.~\ref{fig:Na-MnTe-fit}. The fit quality is good in both cases, indicating that the published structure provides a good model for the data. Equivalent fits were conducted for all data sets collected between 5 and 500~K.

\textbf{Debye-Gr\"uneisen fits.}
The best-fit parameter values for the Debye-Gr\"uneisen fits to the lattice parameters and unit cell volume for pure and doped MnTe are given in Table~\ref{table:params}.
\begin{table}[ht]
	\ra{1.3}
	\caption{Best-fit parameter values from Debye-Gr\"uneisen fits to the lattice parameter data for pure and doped MnTe.} 
	\centering 
	\begin{tabular}{l c c} 
		\hline
        \hline
		& ~~~$V_0$ (\AA$^3$) or $\ell_0$ (\AA)~~~ & $\gamma/B_0$ (Pa$^{-1}$) \\  
		\cline{1-3}
		 \multicolumn{1}{c}{MnTe}& &   \\
        \cline{1-1}
		Volume fits & 98.947(8) & $3.42(1)\times 10^{-11}$  \\
		$a$ parameter fits & 4.137(1) & $2.0(4)\times 10^{-11}$  \\
		$c$ parameter fits & 6.6798(3) & $1.40(1)\times 10^{-10}$  \\
         & & \\
		  \multicolumn{1}{c}{Na-doped MnTe}& &   \\
        \cline{1-1}
		Volume fits & 99.06(1) & $3.35(2)\times 10^{-11}$  \\
		$a$ parameter fits & 4.1358(3) & $2.10(8)\times 10^{-11}$  \\
		$c$ parameter fits & 6.6906(5) & $1.27(2)\times 10^{-10}$  \\
		\hline
        \hline
	\end{tabular}
	\label{table:params} 
\end{table}
The values for $\frac{\gamma}{B_0}$ are similar but not identical for the fits performed against the volume, $a$ lattice parameter, and $c$ lattice parameter. The differences reflect anisotropies in the elastic properties of the solid. Using the volume result for pure MnTe and substituting in the reported bulk modulus of 47.3~GPa~\cite{paszk;appa97}, we can estimate the Gr\"uneisen parameter to be $\gamma = 1.6$, a reasonable value for solids that supports the reliability of this approach. The values of $\gamma/B_0$ are also close to those reported for other binary antiferromagnetic compounds~\cite{chatt;jpcm10, chatt;jpcm11}.

\begin{figure}
	\centerline{\includegraphics[ width=75mm]{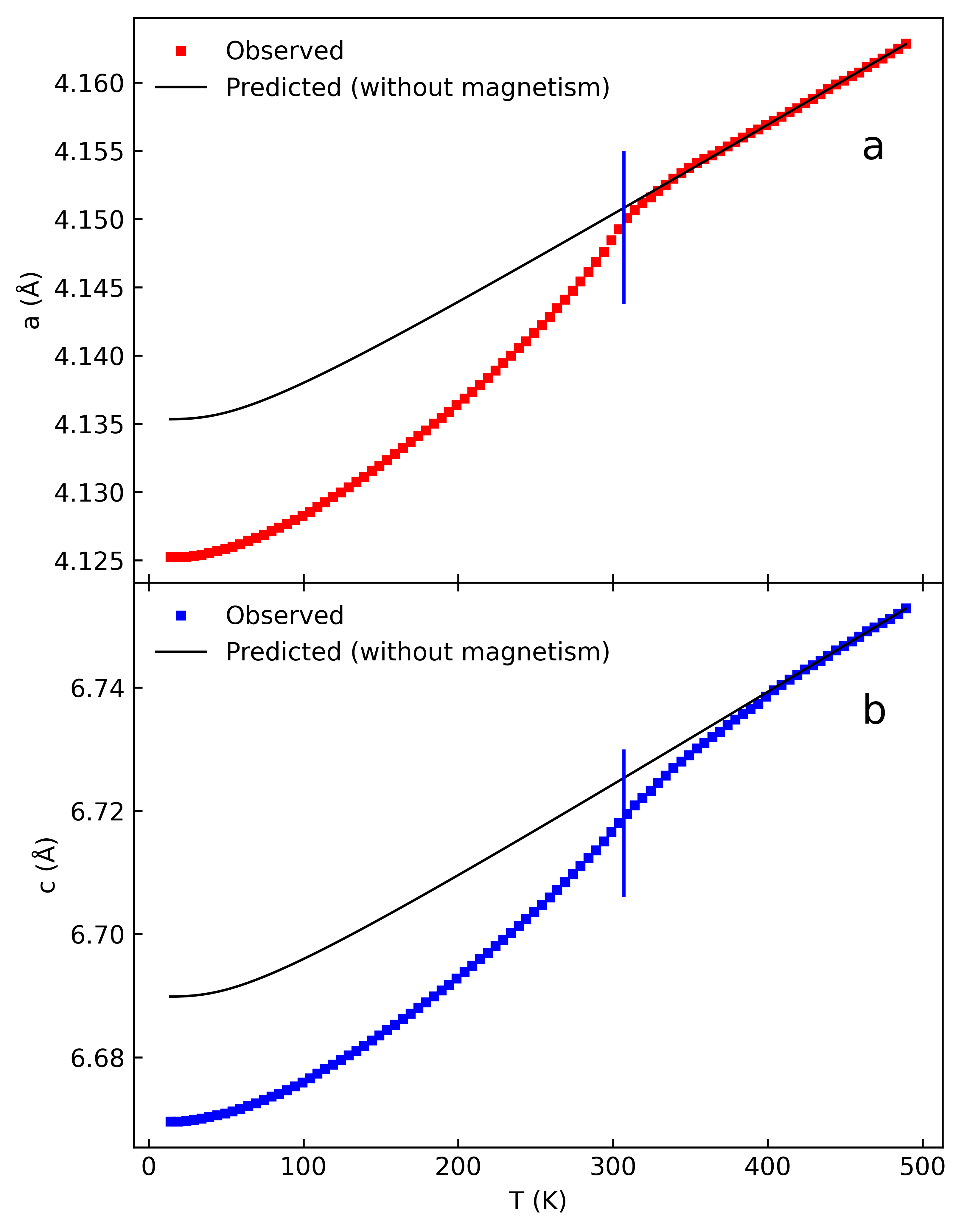}}
	\caption{\label{fig:Na-MnTe-debye} Temperature dependence of the $a$ (panel a) and $c$ (panel b) lattice parameters for Na-doped MnTe determined from x-ray PDF fits. The solid black curve shows the prediction using the Debye-Gr\"uneisen model, which does not take into account magnetostructural effects.}		
\end{figure}

\begin{figure}
	\centerline{\includegraphics[ width=75mm]{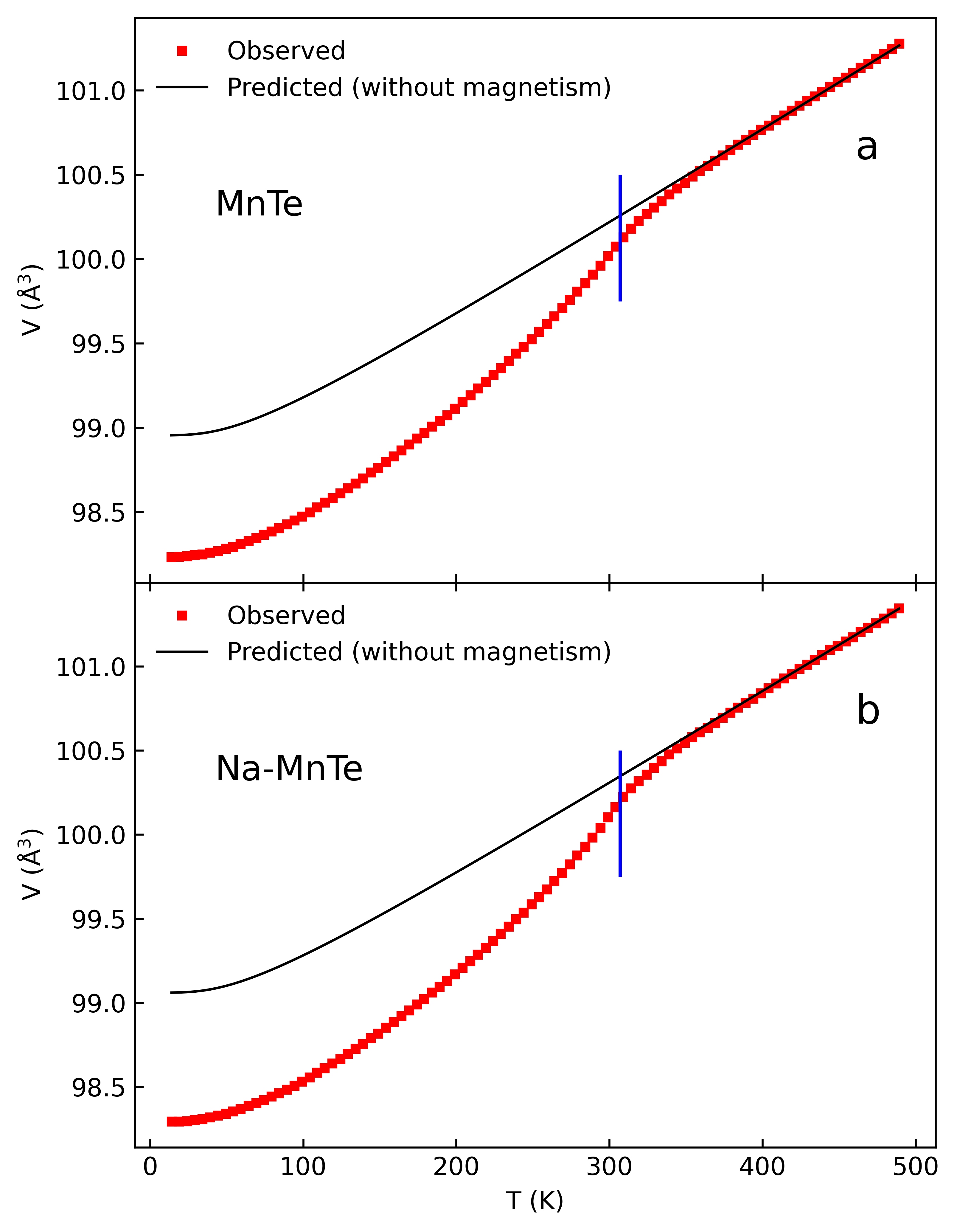}}
	\caption{\label{fig:debye-volume} Temperature dependence of the unit cell volume for pure MnTe (a) and Na-doped MnTe (b) determined from x-ray PDF fits. The solid black curves show the prediction using the Debye-Gr\"uneisen model, which does not take into account magnetostructural effects.}		
\end{figure}

The Debye-Gr\"uneisen fits to the lattice parameters for Na-doped MnTe are shown in Fig.~\ref{fig:Na-MnTe-debye}.
We used a Debye temperature of 223~K, which was also used for pure MnTe. The measured unit cell volume as a function of temperature for both compounds is plotted in Fig.~\ref{fig:debye-volume}, together with the predicted temperature dependence based on the Debye-Gr\"uneisen model.
The pure and doped compounds both show nearly indistinguishable trends. At the lowest temperature, $\Delta V/V$ is $-7.3 \times 10^{-3}$ for pure MnTe and $-7.8 \times 10^{-3}$ for Na-doped MnTe.

\begin{figure}
	\centerline{\includegraphics[ width=150mm]{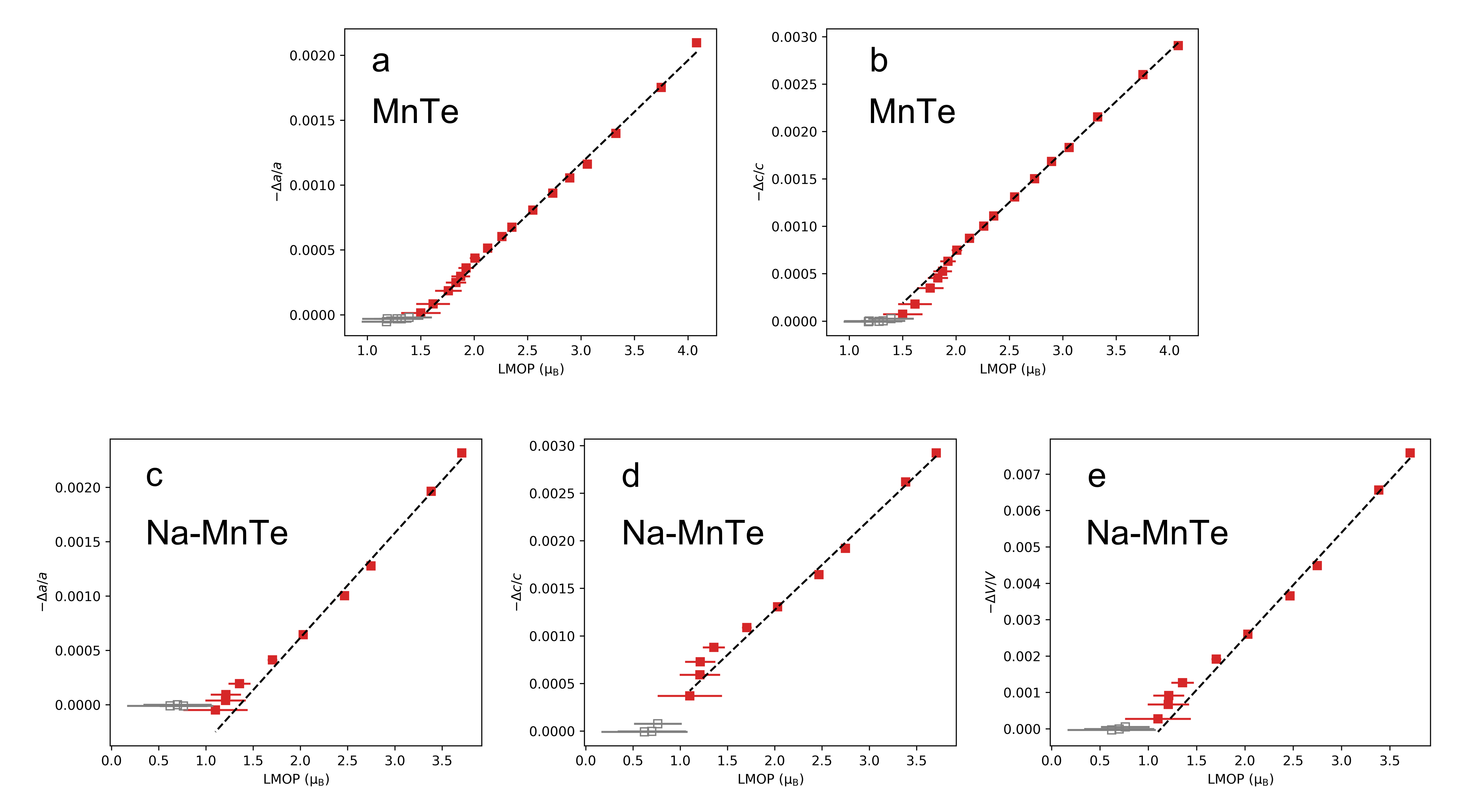}}
	\caption{\label{fig:moment-scaling} (a, b) The fractional change in $a$ (panel a) and $c$ (panel b) versus the LMOP in pure MnTe. The black dashed line is a best-fit line to the red data points. The gray data points were left out of the fit because they represent data collected at high temperatures above the onset of the magnetostructural response. (c-e) Same as (a, b) but for Na-doped MnTe and with the fractional volume shift included as panel (e).}		
\end{figure}

\textbf{Linear scaling of the lattice response with the local magnetic order parameter.} In Fig.~\ref{fig:moment-scaling}, we plot the fractional lattice change versus the local magnetic order parameter (LMOP) for the $a$ and $c$ lattice parameters of pure MnTe as well as $a$, $c$, and the unit cell volume for Na-doped MnTe.
In all cases, a predominantly linear relationship is clearly observed.

\begin{figure}
	\centerline{\includegraphics[ width=180mm]{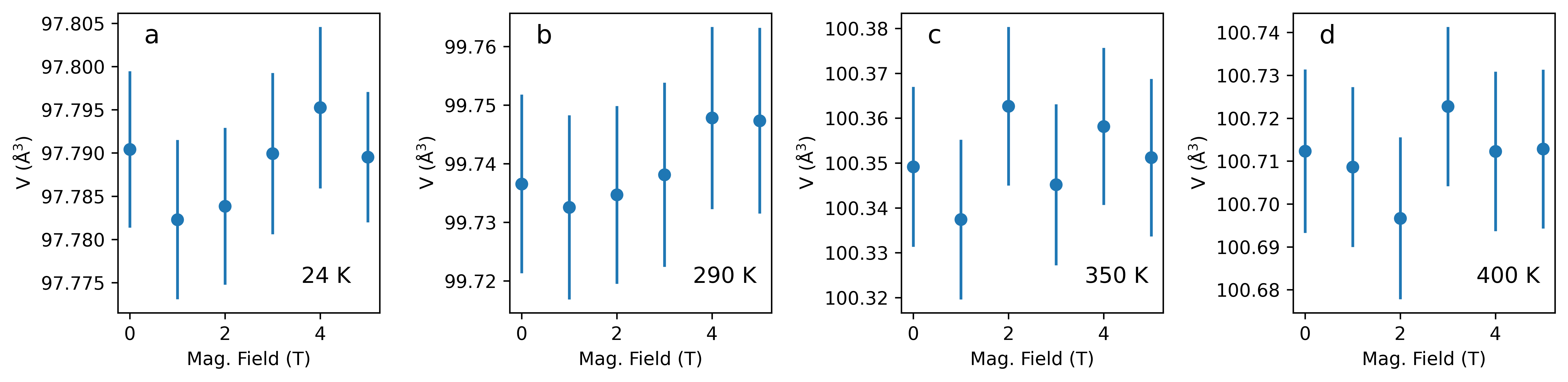}}
	\caption{\label{fig:forced-magnetostriction} Unit cell volume versus applied magnetic field for pure MnTe at 24~K (a), 290~K (b), 350~K (c), and 400~K (d). The error bars represent the estimated standard deviation.}		
\end{figure}

\textbf{Field-dependent PDF measurements.}
To investigate the possibility of forced magnetostriction in MnTe, we performed x-ray PDF measurements with an \textit{in situ} magnetic field up to 5~T. Data were collected for pure MnTe at temperatures of 24~K, 290~K, 350~K, and 400~K. No field-induced change is observed in the individual lattice parameters or the unit cell volume at any temperature, as illustrated in Fig.~\ref{fig:forced-magnetostriction}.
We can estimate a generous upper limit of the forced volumetric magnetostriction in MnTe in fields up to 5~T by considering the difference between the maximum volume plus one estimated standard deviation (ESD) and the minimum volume minus one ESD. The largest upper limit estimated in this way is $\Delta V/V = 6.3 \times 10^{-4}$. An upper limit that is perhaps more realistic can be obtained by fitting a line or parabola to the data points and calculating the largest magnitude of $\Delta V/V$ from the fitted curve. This largest upper limit produced by this procedure is $1.5 \times 10^{-4}$ for 290~K. 

\textbf{Magnetic domain coupling for Na-doped MnTe.}
Lightly doping MnTe with Na or Li causes a reorientation of the spins such that they lie nearly along the $c$ axis at low temperature, gradually canting further away from the $c$ axis toward the in-plane spin arrangement in pure MnTe as the temperature is increased toward the transition~\cite{mosel;prm22,baral;matter22}. Given the small energy difference between the in-plane and out-of-plane spin arrangements, it is reasonable to expect that the short-range correlations in the paramagnetic state may also include out-of-plane components. The out-of-plane spin component corresponds to the one-dimensional $\Gamma_{4}^+$ irreducible representation of space group $P6_3/mmc$ (\#194), which permits two domains corresponding to opposite orientations relative to the $c$ axis. Combining two out-of-plane domains with six in-plane domains results in a set of 12 joint order parameter (OP) domains. Regardless of the number of domains, the fundamental idea is the same, namely that the long-range OP in the reference domain arises at $T_{\mathrm{N}}$, while the short-range OP in other domains pre-exists at $T_{\mathrm{N}}$. When they are coupled, approximately linear scaling arises, as explained in the main text.



%